\documentstyle[emulateapj]{article}

\begin{document}

\title{Untangling the X-ray Emission From the Sa Galaxy NGC~1291 With
{\it Chandra}}

\author{Jimmy A. Irwin\altaffilmark{1,2}, Craig L. Sarazin\altaffilmark{3},
and Joel N. Bregman\altaffilmark{1}}

\altaffiltext{1}{Department of Astronomy, University of Michigan,
Ann Arbor, MI 48109-1090
E-mail: jirwin@astro.lsa.umich.edu, jbregman@umich.edu}

\altaffiltext{2}{Chandra Fellow.}

\altaffiltext{3}{Department of Astronomy, University of Virginia,
P.O. Box 3818, Charlottesville, VA 22903-0818;
E-mail: sarazin@virginia.edu}

\begin{abstract}
We present a {\it Chandra} ACIS-S observation of the nearby bulge-dominated
Sa galaxy NGC~1291. The X-ray emission from the bulge resembles the X-ray
emission from a sub-class of elliptical and S0 galaxies with low
X-ray--to--optical luminosity ratios.
The X-ray emission is composed of a central point-like nucleus, $\sim$50 point
sources that are most likely low mass X-ray binaries (LMXBs), and diffuse gas
detectable out to a radius of 120$^{\prime\prime}$ (5.2 kpc).
The diffuse gas has a global temperature of $0.32^{+0.04}_{-0.03}$ keV and
metallicity of $0.06 \pm 0.02$ solar, and both quantities marginally decrease
with increasing radius.
The hot gas fills the hole in the H~{\sc i}
distribution, and the softening of the spectrum of the X-ray gas with radius
might indicate a thermal coupling of the hot and cold
phases of the interstellar medium as previously suggested. The integrated
X-ray luminosity of the LMXBs, once normalized by the optical luminosity,
is a factor of 1.4 less than in the elliptical galaxy NGC~4697 or
S0 galaxy NGC~1553. The difference
in $L_{X,stellar}/L_B$ between the galaxies appears to be because of a
lack of very bright sources in NGC~1291. No sources above $3 \times 10^{38}$
ergs s$^{-1}$ were found in NGC~1291 when $\sim$7 were expected from
scaling from NGC~4697 and NGC~1553.
The cumulative $L_{X,stellar}/L_B$ value including
only sources below $1.0 \times 10^{38}$ ergs s$^{-1}$ is remarkably
similar between NGC~1291 and NGC~4697, if a recent surface brightness
fluctuation-determined distance is assumed for NGC~4697.
If this is a common feature of the
LMXB population in early-type systems, it might be used as a distance
indicator. Finally, a bright, variable ($1.6-3.1 \times 10^{39}$ ergs s$^{-1}$)
source was detected at the optical center of the galaxy. Its spectrum shows
excess soft emission superimposed on a highly absorbed power law component,
similar to what has been found in several other low luminosity active
galactic nuclei. However,
the soft component does not vary in intensity like the hard component,
indicating that the soft component is not reprocessed hard component emission.
\end{abstract}


\section{Introduction} \label{sec:intro}

 From a stellar population point of view, the large bulges of Sa galaxies
closely resemble elliptical galaxies. It is therefore natural to ask if
this similarity extends into the X-ray regime.
With the launch of {\it Chandra}, our view of the X-ray emission from
elliptical galaxies has become considerably clearer. The nature
of the X-ray emission from X-ray bright (high X-ray--to--optical luminosity
ratio, $L_X/L_B$) elliptical galaxies had been known since the days of the
{\it Einstein Observatory} to be primarily from hot $10^7$ K gas
(e.g., Forman, Jones, \& Tucker 1985; Trinchieri, Fabbiano, \& Canizares 1986),
with a small percentage of the flux from a harder
component (presumably the integrated emission from a collection of
low-mass X-ray binaries, hereafter LMXBs; Matsumoto et al.\ 1997).
However, only {\it Chandra's} $0\farcs5$ spatial resolution was able to resolve
the X-ray emission from X-ray faint (low $L_X/L_B$) elliptical and S0 galaxies
into its constituent parts. In the X-ray faint elliptical galaxy NGC~4697
{\it Chandra} found that a majority of the X-ray emission emanated from a
collection of point sources, primarily LMXBs, with the remaining X-ray
emission from a lower temperature diffuse interstellar medium (Sarazin,
Irwin, \& Bregman 2000; hereafter SIB00).
In these galaxies, much of the hot interstellar medium
has apparently been removed, leaving behind the X-ray
binaries as the primary X-ray emission mechanism.

Whether spiral bulges more closely resemble the gas-rich X-ray bright
elliptical galaxies or the stellar-dominated X-ray faint galaxies
remains undetermined. Previous work with {\it Einstein}
indicated that the average X-ray temperature of Sa galaxies was intermediate
between elliptical galaxies and later-type spiral galaxies
(Kim, Fabbiano, \& Trinchieri 1992). Further comparisons with {\it ROSAT}
revealed that the bulges of the two optically brightest Sa galaxies (NGC~1291
and NGC~3623) and the bulge of M31 had similar $L_X/L_B$ values and X-ray
colors as X-ray faint elliptical galaxies (Irwin \& Sarazin 1998a,b).
Recent {\it ROSAT} HRI observations of NGC~1291 found that the radial surface
brightness distribution of the X-rays followed a de Vaucouleurs profile
suggesting that stellar X-ray sources were responsible for the observed X-ray
emission (Hogg et al.\ 2001), although {\it ROSAT} PSPC data indicated that
X-ray spectrum softened with increasing radius, which would not be expected
from a purely stellar component (Bregman, Hogg, \& Roberts 1995).
But whether the X-ray emission from spiral bulges is dominated by a stellar
component can only be confirmed using the excellent spatial resolution of
{\it Chandra}. Here, we present a {\it Chandra} observation of the
nearby Sa galaxy NGC~1291 to resolve the question of the X-ray emission
in spiral galaxies with large bulges.

\section{Target and Data Reduction} \label{sec:data}

NGC~1291 is the optically brightest Sa galaxy in the Revised Shapley Ames
Catalog with a corrected apparent magnitude of $B_0=9.17$. Detailed optical
imaging by de Vaucouleurs (1975) revealed a central lens surrounded by a large
outer ring from which two faint spiral arms emanate. Sometimes classified as
a transitional SB(s)0/Sa galaxy, there are no other optically bright
galaxies in velocity space near NGC~1291, although recently a large
concentration of faint low surface brightness galaxies have been found
around NGC~1291 (Kambas et al.\ 2000). The distance to NGC~1291
is uncertain, and literature values range from 6.9 Mpc to 13.8 Mpc.
de Vaucouleurs (1975) gives a distance of 8.9 Mpc by assuming that the
brightest stars in the weak spiral arms of NGC~1291 have an absolute
magnitude of $M_B=-9.5$, and this is the distance we have assumed for this
work. The corrected galactocentric velocity is 674 km s$^{-1}$
(de Vaucouleurs 1975). At a distance
of 8.9 Mpc, 1$^{\prime}$ corresponds to 2.6 kpc.

In this paper, we will concentrate on X-ray sources and diffuse emission
seen in projection within the optical bulge of NGC~1291.
We will define the bulge to be the region within an ellipse centered on
the optical nucleus,
with major and minor axes of 4\farcm5  and 3\farcm7, respectively, and
at a position angle of $170^{\circ}$ (de Vaucouleurs 1975).

NGC~1291 was observed with the Advanced CCD Imaging Spectrometer (ACIS)
onboard {\it Chandra} at two epochs, on 2000 June 27--28,
and again on 2000 November 7. The first observation (hereafter OB1) was
awarded to us in AO-1 of {\it Chandra}, while the second observation
\centerline{\null}
\vskip3.10truein
\includegraphics{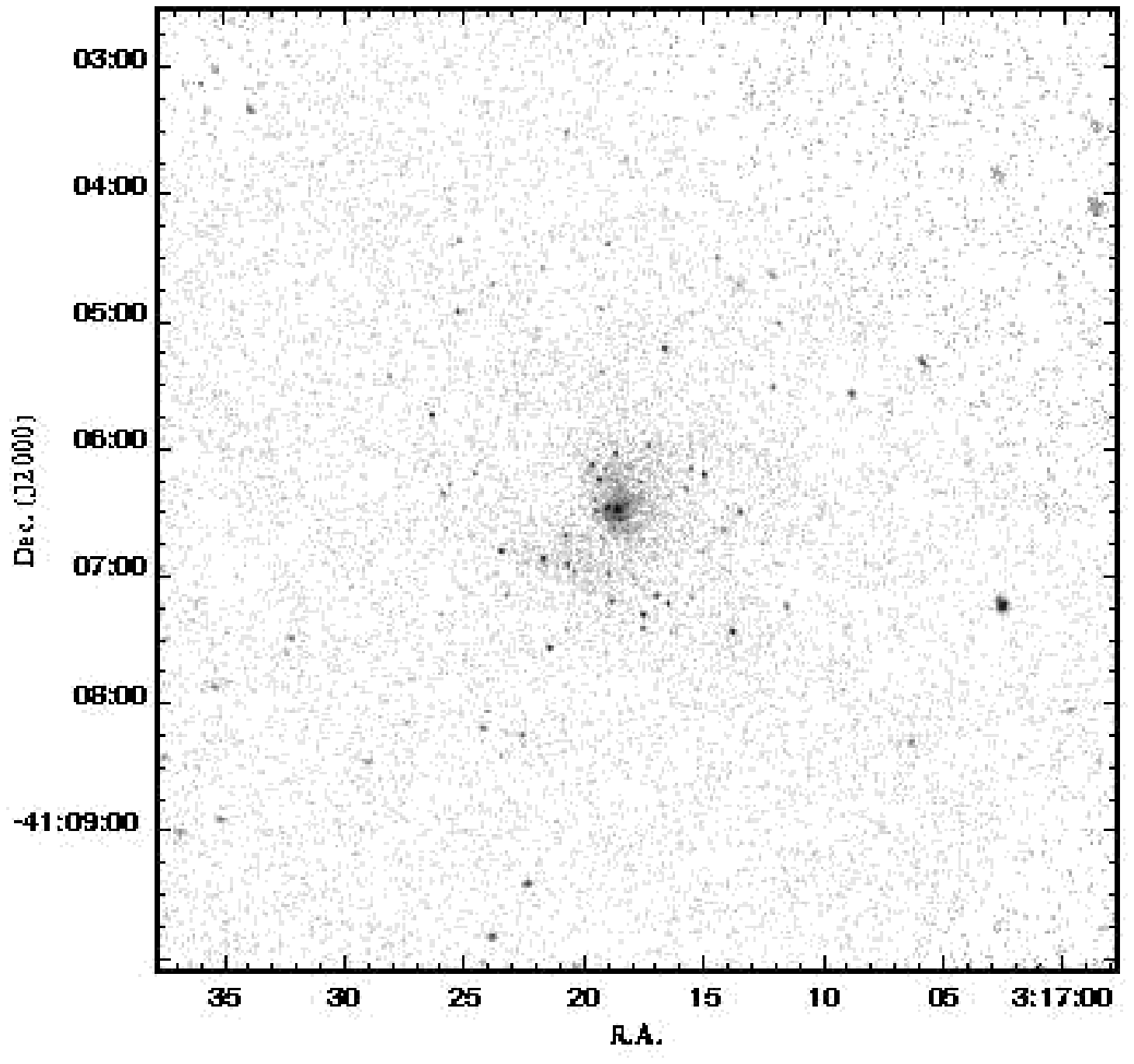}
\figcaption{ \small
0.3--6.0 keV image of the OB1+OB2 ACIS-S observation of NGC~1291. The image has
been smoothed with a Gaussian of width $0\farcs5$ to bring out point sources and
diffuse emission.
\label{fig:xray}}
\normalsize
(hereafter OB2) was graciously donated to the {\it Chandra} archive
early by Andrea Prestwich. OB1 contained one small background flare, while
over a third of OB2 was taken during times of high background. The total
usable exposure for OB1 and OB2 was 37,409 seconds and 23,016 seconds,
respectively. Both OB1 and OB2 were taken when the focal plane temperature was
--120$^\circ$ C. The observations were combined for the spatial analysis, and
treated separately for the spectral analysis. We present data only from
the backside-illuminated S3 chip. Exposure maps were generated to correct
the images for vignetting. An exposure map appropriate for an energy of
0.75 keV was generated for soft energy images, while an exposure map
appropriate for an energy of 3 keV was created for hard energy images.
The newest spectral responses available (acisD2000-01-29fef\_piN0002.fits)
were used in the spectral analysis.

\section{Spatial Analysis} \label{sec:spatial}

\subsection{X-ray Images} \label{ssec:images}

The 0.3--6 keV {\it Chandra} image of the merged OB1+OB2
observation is shown in Figure~\ref{fig:xray}. It was found that omitting
energy channels above 6.0 keV reduced the background by 51\% while decreasing
the source emission by only 4\%. The image has been lightly smoothed
\centerline{\null}
\vskip5.60truein
\includegraphics{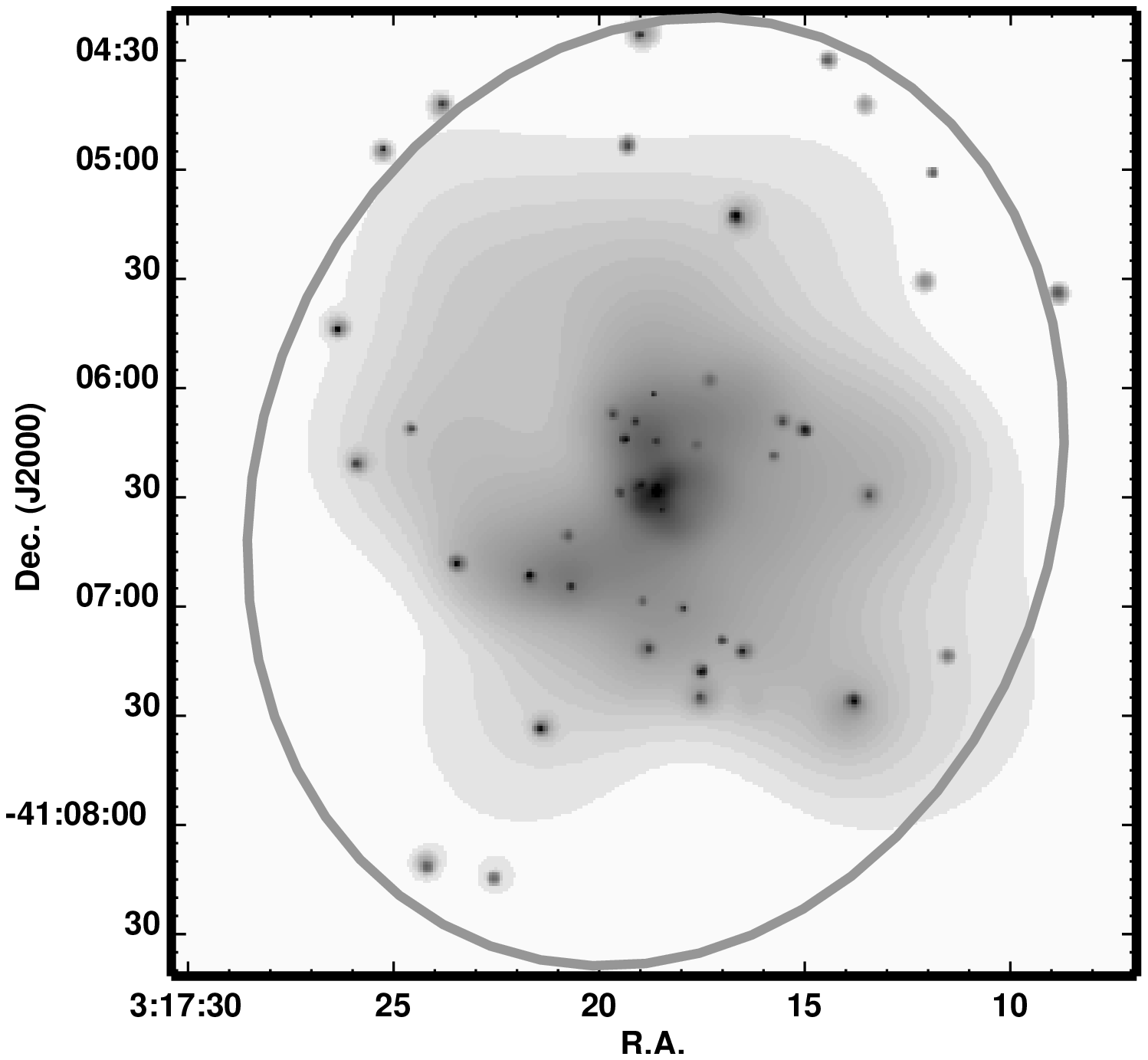}
\includegraphics{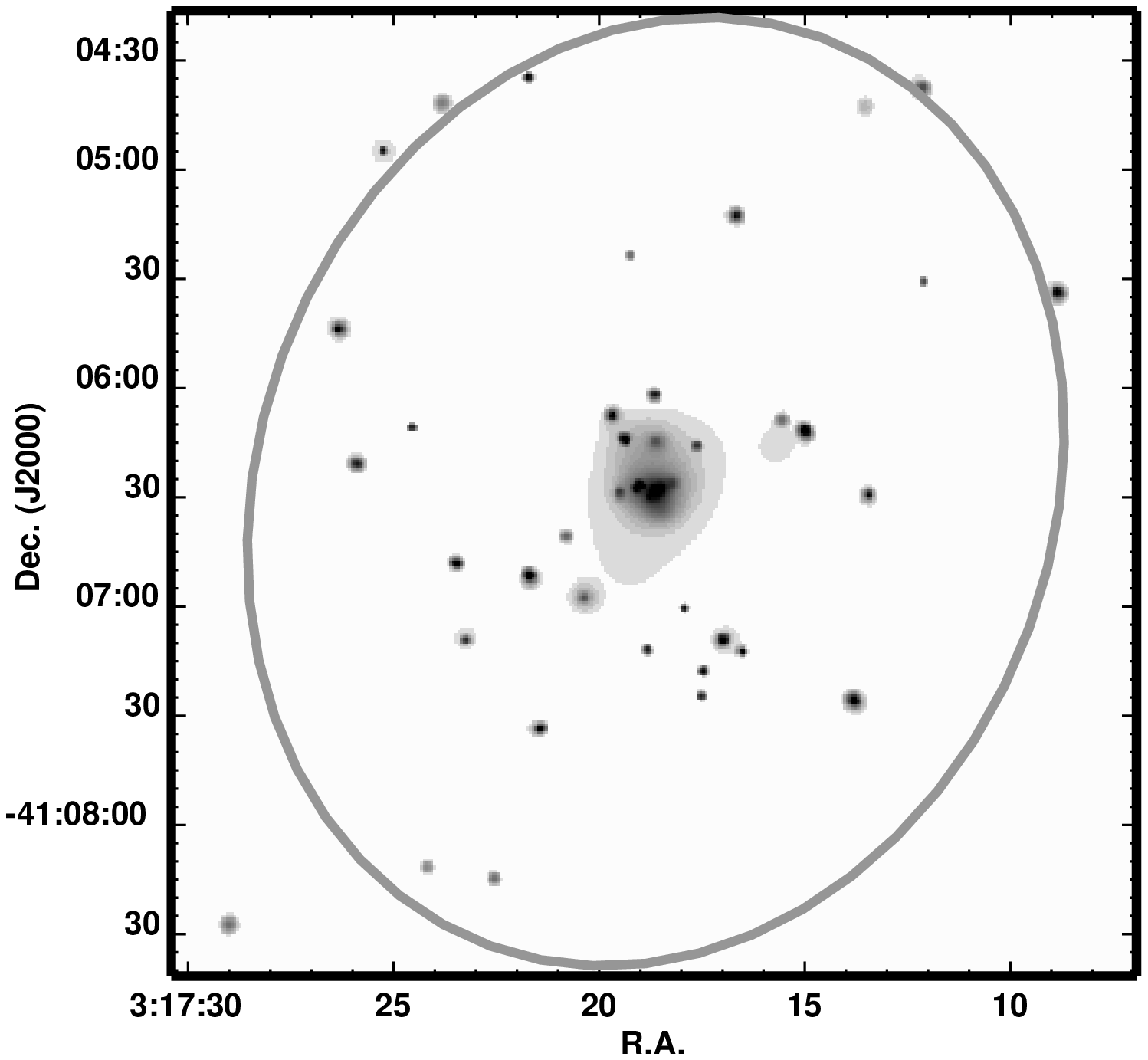}
\figcaption{\small
Inner $4\farcm5 \times 4\farcm5$ of NGC~1291 in the $(a)$ 0.3-1.0 keV
({\it upper}) and $(b)$ 1.5-6.0 keV ({\it lower}) energy band.
The images have been adaptively smoothed
to a minimum signal--to--noise ratio of three per smoothing beam. The
ellipse represents the optical extent of the bulge as determined by
de Vaucouleurs (1975).
\label{fig:adapt}}

\begin{table*}[t]
\footnotesize
\begin{center}
\caption{Discrete X-ray Sources in Bulge \label{tab:sources}}
\begin{tabular}{lccccccc}
\tableline
\tableline
Src.&Name& R.A.& Dec.& $d$ &  Count Rate& SNR & $L_X$ (0.3-10 keV)\cr
No..&&(h:m:s)& ($\arcdeg$:$\arcmin$:$\arcsec$)& (arcsec) &
($10^{-4}$ s$^{-1}$)& & ($10^{37}$ ergs s$^{-1}$) \cr
(1)&(2)&(3)&(4)&(5)&(6)&(7)&(8)\cr
\tableline
 1&CXOU J031718.6-410628& 3:17:18.60&$-41$:06:28.3&\phn\phn0.00&255.25$\pm$9.09& 28.09&305.20\cr
 2&CXOU J031718.7-410630& 3:17:18.79&$-41$:06:30.2&\phn\phn2.86&\phn 19.67$\pm$2.52&\phn  7.80&\phn 13.39\cr
 3&CXOU J031718.9-410627& 3:17:18.98&$-41$:06:27.0&\phn\phn4.52& \phn25.50$\pm$2.87& \phn 8.88&\phn 17.36\cr
 4&CXOU J031718.2-410626& 3:17:18.22&$-41$:06:26.4&\phn\phn4.70& \phn\phn8.96$\pm$ 1.70& \phn 5.26&\phn\phn  6.10\cr
 5&CXOU J031718.4-410633& 3:17:18.48&$-41$:06:33.4&\phn\phn5.21& \phn11.49$\pm$1.93& \phn 5.96&\phn\phn  7.82\cr
 6&CXOU J031719.1-410628& 3:17:19.12&$-41$:06:28.1&\phn\phn5.92& \phn19.71$\pm$2.53& \phn 7.81&\phn 13.42\cr
 7&CXOU J031718.7-410637& 3:17:18.75&$-41$:06:37.3&\phn\phn9.12& \phn\phn 5.33$\pm$1.31& \phn 4.06&\phn\phn  3.63\cr
 8&CXOU J031719.4-410629& 3:17:19.47&$-41$:06:29.1&\phn\phn9.95& \phn\phn 6.69$\pm$1.47& \phn 4.55&\phn\phn  4.56\cr
 9&CXOU J031719.5-410623& 3:17:19.55&$-41$:06:24.0&\phn 11.57& \phn\phn 5.62$\pm$1.35& \phn 4.17&\phn\phn  3.83\cr
10&CXOU J031718.6-410615& 3:17:18.63&$-41$:06:15.1&\phn 13.22& \phn\phn 8.23$\pm$1.63& \phn 5.05&\phn\phn  5.61\cr
11&CXOU J031719.2-410641& 3:17:19.28&$-41$:06:41.9&\phn 15.63& \phn\phn 4.49$\pm$1.20& \phn 3.73&\phn\phn  3.06\cr
12&CXOU J031717.6-410615& 3:17:17.65&$-41$:06:15.9&\phn 16.38& \phn\phn 7.61$\pm$1.57& \phn 4.85&\phn\phn  5.18\cr
13&CXOU J031719.3-410614& 3:17:19.37&$-41$:06:14.4&\phn 16.42& \phn34.04$\pm$3.32& 10.26&\phn 23.18\cr
14&CXOU J031719.1-410609& 3:17:19.13&$-41$:06:09.4&\phn 19.84& \phn\phn 4.89$\pm$1.26& \phn 3.89&\phn\phn  3.33\cr
15&CXOU J031719.6-410607& 3:17:19.68&$-41$:06:07.7&\phn 23.96& \phn12.21$\pm$1.99& \phn 6.14&\phn\phn  8.31\cr
16&CXOU J031718.6-410602& 3:17:18.66&$-41$:06:02.2&\phn 26.08& \phn\phn 9.76$\pm$1.78& \phn 5.49&\phn\phn  6.64\cr
17&CXOU J031720.7-410641& 3:17:20.79&$-41$:06:41.0&\phn 27.85& \phn\phn 4.54$\pm$1.21& \phn 3.75&\phn\phn  3.09\cr
18&CXOU J031718.9-410658& 3:17:18.96&$-41$:06:58.7&\phn 30.67& \phn\phn 3.26$\pm$1.03& \phn 3.18&\phn\phn  2.22\cr
19&CXOU J031717.9-410700& 3:17:17.95&$-41$:07:00.7&\phn 33.22& \phn\phn 8.33$\pm$1.64& \phn 5.08&\phn\phn  5.67\cr
20&CXOU J031717.3-410558& 3:17:17.30&$-41$:05:58.4&\phn 33.28& \phn\phn 5.85$\pm$1.38& \phn 4.25&\phn\phn  3.98\cr
21&CXOU J031720.3-410657& 3:17:20.39&$-41$:06:58.0&\phn 35.90& \phn\phn 4.88$\pm$1.26& \phn 3.88&\phn\phn  3.32\cr
22&CXOU J031715.5-410609& 3:17:15.56&$-41$:06:09.3&\phn 39.23& \phn\phn 7.33$\pm$1.54& \phn 4.76&\phn\phn  4.99\cr
23&CXOU J031721.7-410651& 3:17:21.70&$-41$:06:51.9&\phn 42.24& \phn32.59$\pm$3.25& 10.04&\phn 22.19\cr
24&CXOU J031718.8-410712& 3:17:18.82&$-41$:07:12.0&\phn 43.77& \phn11.93$\pm$1.96& \phn 6.07&\phn\phn  8.12\cr
26&CXOU J031716.9-410709& 3:17:16.99&$-41$:07:09.5&\phn 45.02& \phn12.49$\pm$2.01& \phn 6.21&\phn\phn  8.50\cr
27&CXOU J031716.5-410712& 3:17:16.52&$-41$:07:12.7&\phn 50.21& \phn15.37$\pm$2.23& \phn 6.89&\phn 10.47\cr
28&CXOU J031717.5-410718& 3:17:17.51&$-41$:07:18.2&\phn 51.42& \phn19.85$\pm$2.53& \phn 7.83&\phn 13.52\cr
29&CXOU J031717.5-410724& 3:17:17.53&$-41$:07:24.9&\phn 57.81& \phn\phn 6.40$\pm$1.44& \phn 4.45&\phn\phn  4.36\cr
30&CXOU J031713.4-410629& 3:17:13.46&$-41$:06:29.6&\phn 58.04& \phn\phn 9.92$\pm$1.79& \phn 5.54&\phn\phn  6.75\cr
31&CXOU J031723.4-410648& 3:17:23.48&$-41$:06:48.6&\phn 58.73& \phn21.63$\pm$2.64& \phn 8.18&\phn 14.73\cr
32&CXOU J031723.2-410709& 3:17:23.25&$-41$:07:09.4&\phn 66.75& \phn\phn 4.24$\pm$1.17& \phn 3.62&\phn\phn  2.89\cr
33&CXOU J031724.5-410611& 3:17:24.57&$-41$:06:11.2&\phn 69.67& \phn\phn 6.63$\pm$1.46& \phn 4.53&\phn\phn  4.52\cr
34&CXOU J031721.4-410733& 3:17:21.43&$-41$:07:33.9&\phn 73.03& \phn22.39$\pm$2.69& \phn 8.32&\phn 15.24\cr
35&CXOU J031716.6-410512& 3:17:16.68&$-41$:05:12.9&\phn 78.46& \phn36.23$\pm$3.42& 10.58&\phn 24.67\cr
36&CXOU J031713.8-410726& 3:17:13.82&$-41$:07:26.1&\phn 79.09& \phn34.36$\pm$3.33& 10.31&\phn 23.40\cr
37&CXOU J031725.9-410621& 3:17:25.90&$-41$:06:21.2&\phn 82.86& \phn11.33$\pm$1.91& \phn 5.92&\phn\phn  7.72\cr
38&CXOU J031711.5-410714& 3:17:11.57&$-41$:07:14.1&\phn 91.65& \phn\phn 5.27$\pm$1.31& \phn 4.04&\phn\phn  3.59\cr
39&CXOU J031712.1-410531& 3:17:12.10&$-41$:05:31.3&\phn 92.94& \phn\phn 6.53$\pm$1.45& \phn 4.49&\phn\phn  4.45\cr
40&CXOU J031719.2-410453& 3:17:19.29&$-41$:04:53.7&\phn 94.94& \phn\phn 4.83$\pm$1.25& \phn 3.87&\phn\phn  3.29\cr
41&CXOU J031726.3-410544& 3:17:26.36&$-41$:05:44.1&\phn 98.22& \phn20.86$\pm$2.60& \phn 8.03&\phn 14.21\cr
42&CXOU J031711.8-410501& 3:17:11.89&$-41$:05:01.1&115.55& \phn\phn 3.06$\pm$1.00& \phn 3.08&\phn\phn  2.09\cr
43&CXOU J031722.5-410815& 3:17:22.58&$-41$:08:15.2&115.94& \phn\phn 6.67$\pm$1.47& \phn 4.54&\phn\phn  4.54\cr
44&CXOU J031721.7-410434& 3:17:21.71&$-41$:04:34.9&118.76& \phn\phn 4.71$\pm$1.23& \phn 3.82&\phn\phn  3.21\cr
45&CXOU J031713.5-410442& 3:17:13.54&$-41$:04:42.5&120.22& \phn\phn 4.13$\pm$1.16& \phn 3.57&\phn\phn  2.81\cr
46&CXOU J031724.2-410811& 3:17:24.21&$-41$:08:11.7&121.31& \phn\phn 8.43$\pm$1.65& \phn 5.11&\phn\phn  5.74\cr
47&CXOU J031719.0-410423& 3:17:19.01&$-41$:04:23.2&125.15& \phn 10.41$\pm$1.83& \phn 5.67&\phn\phn  7.09\cr
48&CXOU J031714.4-410430& 3:17:14.45&$-41$:04:30.2&127.12& \phn\phn 3.73$\pm$1.10& \phn 3.40&\phn\phn  2.54\cr
\tableline
\end{tabular}
\end{center}
\end{table*}
with a Gaussian with a width of one pixel to bring out the point sources.
Adaptively-smoothed
images of the inner $4\farcm5 \times 4\farcm5$ for two energy bands
(0.3--1.0 keV and 1.5--6.0 keV) are shown in
Figure~\ref{fig:adapt}.
The images illustrate the major X-ray-emitting
components of the galaxy: a very bright central point source (most likely
an active galactic nucleus), many fainter point sources (most of which are
LMXBs in NGC~1291), and centrally-concentrated diffuse emission that is much
more pronounced in the soft band. The diffuse emission is more or less
circularly symmetric, although there are some asymmetries at intermediate
radii.
These asymmetries argue that the origin of most of the soft emission is
hot gas rather than from a stellar component (which should be distributed like
the optical light). It is unlikely that the majority of the asymmetric
diffuse emission is
from unresolved sources below the detection limit, since similar diffuse
features are not seen in the 1.5--6.0 keV image. Nor is it likely that sources
below the detection threshold of this observation have extremely soft spectra,
since our preliminary work on sources in the bulge of M31 shows that there
is no softening of LMXB spectra for sources with luminosities down to
$10^{36}$ ergs s$^{-1}$.
The gas is contained almost entirely within the optical
confines of the bulge, which is shown in Figure~\ref{fig:adapt}.

\subsection{Point Sources} \label{ssec:sources}

We used the {\sc ciao} program {\sc wavdetect} to detect X-ray point sources
in OB1 over the 0.3-6.0 keV energy range. A minimum detection
threshold of 3$\sigma$ was adopted. This corresponded to a limiting
0.3--6.0 keV count flux of $3.0 \times 10^{-4}$ counts s$^{-1}$ and a
limiting 0.3--10 keV luminosity
of $2.0 \times 10^{37}$ ergs s$^{-1}$ at the assumed distance of 8.9 Mpc,
assuming the spectra of all sources but the central source are described
by a power law with $\Gamma=1.56$ (see \S~\ref{ssec:sources_spec} below).
Here, we concentrate on the sources seen in projection within the bulge
of NGC~1291.
A total of 48 sources were detected from the bulge region in OB1,
and are listed in
Table~\ref{tab:sources} in order of increasing distance from the center
of NGC~1291. For the count--to--luminosity conversion, we assume
1 count s$^{-1} = 6.80 \times 10^{40}$ ergs s$^{-1}$ (0.3--10 keV band)
except for the central
source, where we assume a spectrum based on the spectral fit for that source
(\S~\ref{ssec:central_spec}).

We constructed the luminosity function of the bulge X-ray sources.
Since the luminosity of LMXBs in our own Galaxy can vary on timescales shorter
than a few months, it is unclear whether merging
the two observations will lead to an accurate assessment of the luminosity
function of the sources in NGC~1291. Ideally one wants the
instantaneous distribution of the luminosities, so we analyze only sources
detected in the longer of the two observations, OB1 (although it should be
noted that the luminosity function of OB1+OB2 was not
substantially different from that of OB1 alone). The cumulative
luminosity function for the 47 non-central sources detected in OB1 is
shown in Figure~\ref{fig:xlf}. The central source was not included,
as it is most likely an AGN.
The luminosity function is roughly a power-law between
$3 \times 10^{37}$ ergs s$^{-1}$ and $2 \times 10^{38}$ ergs s$^{-1}$.
There is a steep drop off above $\sim$$2.5 \times 10^{38}$ ergs s$^{-1}$.
The luminosity function also may flatten somewhat below
$\sim$$3 \times 10^{37}$ ergs s$^{-1}$.
However, it is possible that this is due to incompleteness near the
detection threshold ($2.0 \times 10^{37}$ ergs s$^{-1}$), which might be
due to the increase in the width of the instrumental point-spread
function with distance
from the center, or to reduced point-source sensitivity in regions
of high diffuse surface brightness.
Simulations suggest that incompleteness affects the luminosity function
up to a luminosity of about 1.5 times the detection threshold
(e.g., Blanton, Sarazin, \& Irwin 2001).
Thus, we will only fit the observed luminosity function for
$L_X \ge 3 \times 10^{37}$ ergs s$^{-1}$.
\centerline{\null}
\vskip2.50truein
\includegraphics{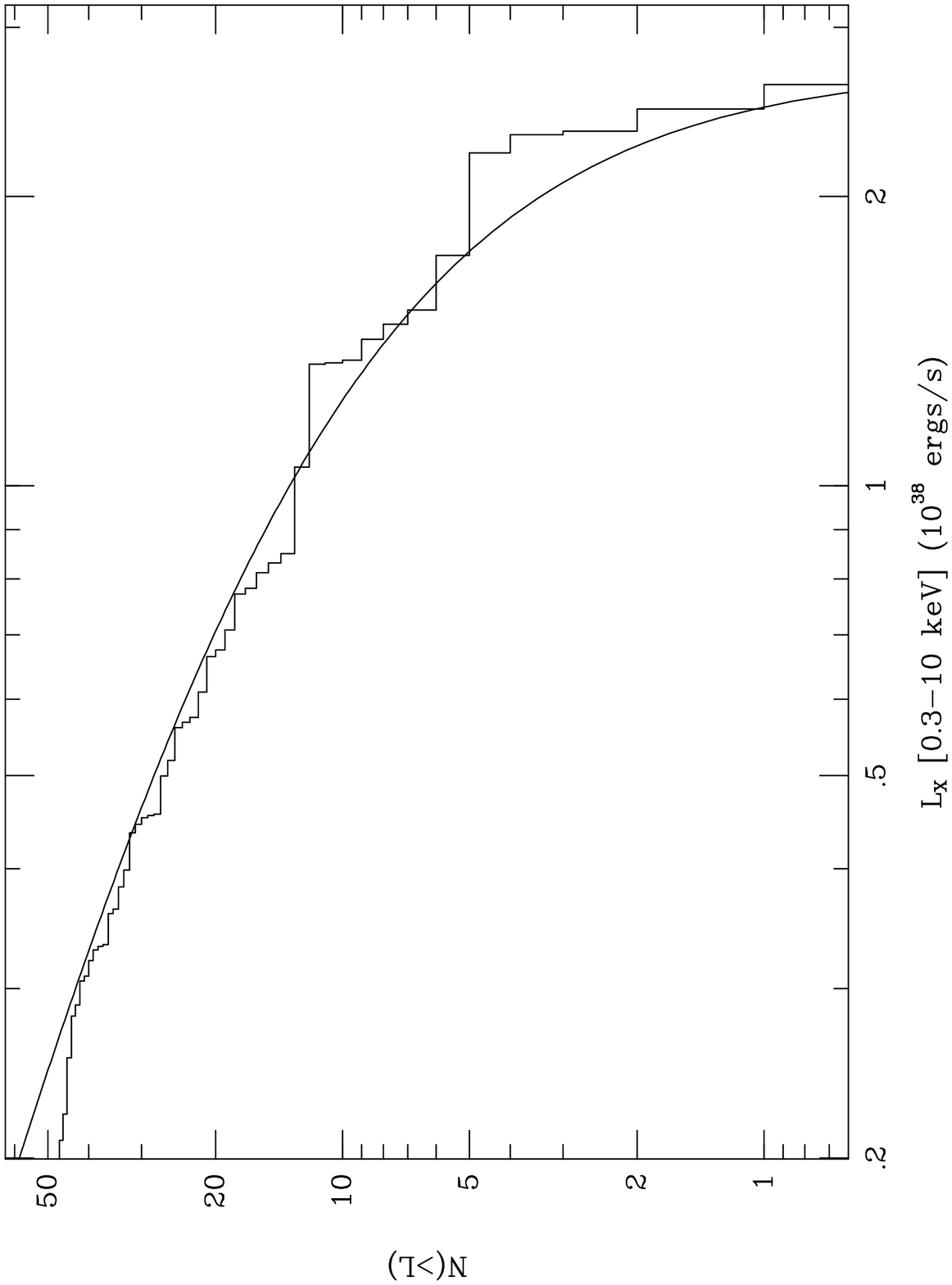}
\figcaption{\small
Histogram of the cumulative 0.3--10 keV X-ray luminosity function
for the 47 point sources within the bulge of NGC~1291 detected in OB1.
The central source has been omitted.
Roughly three of these sources are expected to be unrelated
foreground/background
sources.
Incompleteness may affect the sources below $\sim$$3 \times 10^{37}$ ergs
s$^{-1}$.
The solid line is the best-fit cut-off power-law function
for the galactic sources
(eq.~\protect\ref{eq:xlum}).
\label{fig:xlf}}

We used maximum likelihood techniques to fit the luminosity function,
as described in
SIB00 and
Sarazin, Irwin, \& Bregman (2001, hereafter SIB01).
We modeled unrelated background sources based on
{\it Chandra} deep fields
(e.g., Mushotzky et al.\ 2000);
about three such sources are expected in angular area subtended by the bulge.
We first tried a single power-law (plus the background sources) model,
but it could be rejected at $>$99\% confidence level.
The problem with a single power-law is the steep drop above
$\sim$$2.5 \times 10^{38}$ ergs s$^{-1}$.
Elliptical and S0 galaxies have luminosity functions that are broken
power-laws, with the break occurring at about this same luminosity
(SIB00;
Blanton et al.\ 2001).
Thus, we tried such a broken power-law, and it gave an acceptable fit.
However, the power-law exponent above the break was very steep;
best-fit exponent was $-59$.
Thus, the NGC~1291 source luminosity function effectively is cut-off
above $\sim$$2.5 \times 10^{38}$ ergs s$^{-1}$.
We tried fitting the luminosity function as a single power-law
with a high luminosity cut-off,
\begin{equation} \label{eq:xlum}
\frac{ d N }{ d L_{38} } = \left\{
\begin{array}{lr}
N_o \, \left( \frac{ L_{38} }{ L_c } \right)^{-\alpha} & L_{38} \le L_c \\
0 & L_{38} > L_c \\
\end{array} \right. \, ,
\end{equation}
where $L_{38}$ is the X-ray luminosity (0.3--10 keV) in units
of $10^{38}$ ergs s$^{-1}$, and $L_c$ is the cut-off luminosity.
The best-fit values are $L_c = 2.6^{+0.2}_{-0.5} \times 10^{38}$
ergs s$^{-1}$,
$\alpha = 1.56^{+0.32}_{-0.53}$,
and
$N_o = 3.7^{+4.4}_{-0.9}$.
This best-fit cut-off power-law is shown in Figure~\ref{fig:xlf}.

\centerline{\null}
\vskip3.30truein
\includegraphics{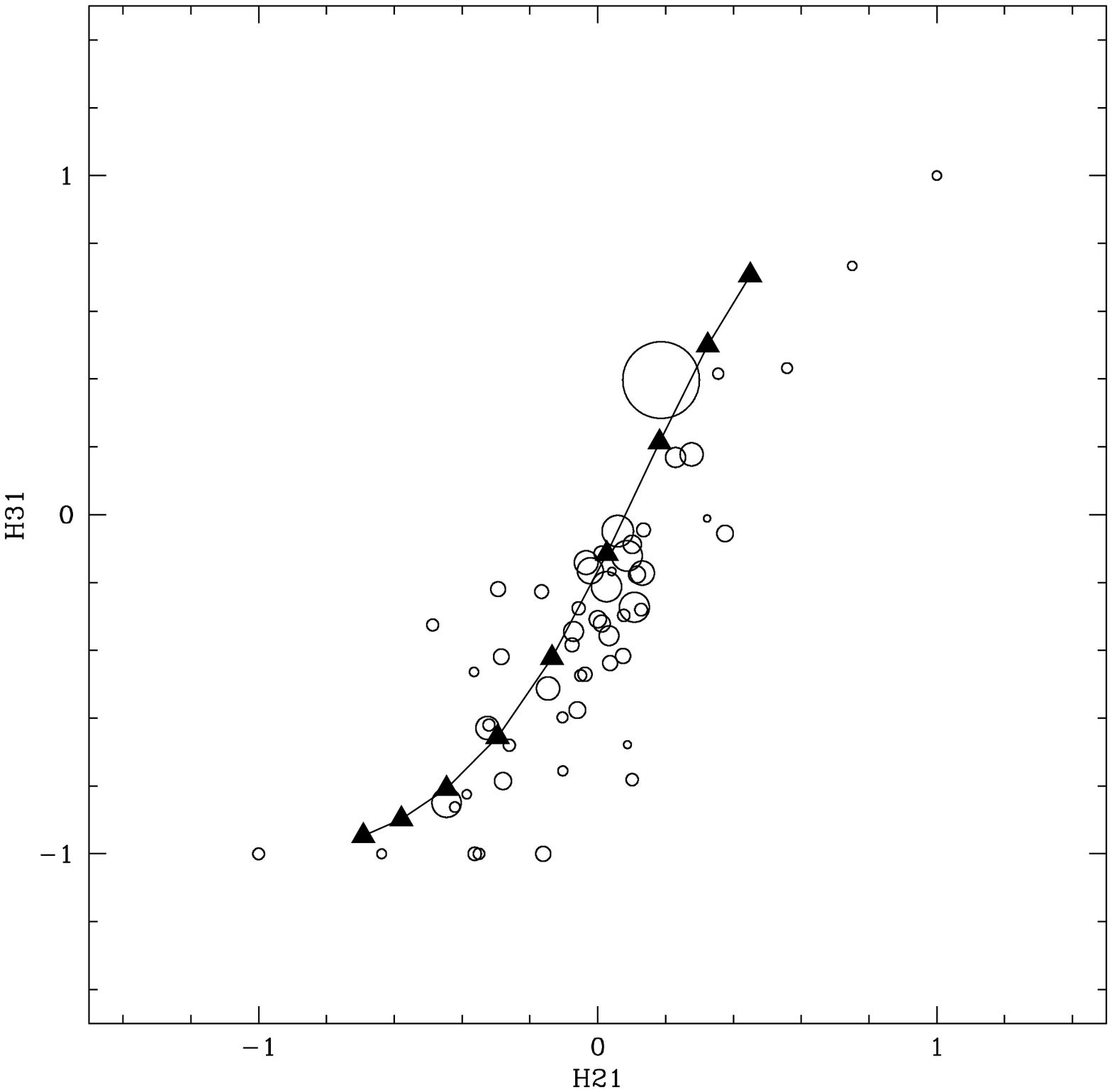}
\figcaption{\small
X-ray hardness ratios of point sources in NGC~1291, defined as
H21=(M--S)/(M+S) and H31=(H--S)/(H+S), where S, M, and H are the counts in the
0.3--1.0 keV, 1.0--2.0 keV, 2.0--6.0 keV energy band, respectively.
The area of the circle is proportional to the total flux of the source.
The filled triangles represent the colors predicted from a power law model with
an exponent $\Gamma=0$ (upper right) to $\Gamma=3.2$ in increments of 0.4.
\label{fig:colors}}
In Figure~\ref{fig:colors} we plot the X-ray colors of the point sources
found in the OB1+OB2 observation for the 48 sources found within the
bulge.
We define three X-ray bands, S (0.3--1.0 keV),
M (1.0--2.0 keV), and H (2.0--6.0 keV), and create two X-ray colors,
H21 = (M--S)/(M+S) and H32 = (H--S)/(H+S). These are the same bands and
colors defined in SIB00, except for the upper energy limit
of band H, where we exclude energies above 6.0 keV to reduce noise. This has
only a minimal effect on H31. The sources occupy a rather narrow band in color
space, as was found for the sources in NGC~4697 and NGC~1553 (SIB01;
Blanton et al.\ 2001). The source located at $(-1,-1)$ is a
supersoft source, similar to those found in the Large Magellanic Cloud and M31
(e.g., Kahabka \& van den Heuvel 1997), and has no detectable emission
above 1 keV. The colors predicted from a power law with a range of exponents
is also shown in Figure~\ref{fig:colors}. The colors of the central source
(largest circle) indicates a rather hard spectrum, although somewhat
more band S emission than might be expected. A detailed spectrum of this
source is presented in \S~\ref{ssec:central_spec} below.
Four sources have H31 values
of --1 (no band H emission) and H21 between --0.2 and --0.4. Three of
these sources are on the outskirts of the bulge and might therefore be unrelated
to NGC~1291 (a similar population of sources was found at larger radii
in NGC~4697; SIB01).

We compared the positions of the X-ray point sources to objects found on the
Digital Sky Survey (DSS) images. None of the 47 non-central point sources
in the bulge of NGC~1291 have optical counterparts that are detectable
on the DSS images. However, this is not surprising since faint optical sources
would be difficult to detect in the high optical surface brightness bulge.
Of the 43 sources outside the bulge, only three had
optical counterparts within $3^{\prime\prime}$,
whose coordinates were obtained from the USNO-A2.0 optical catalog
(Monet et al.\ 1998).
All three optical sources
were very red, and might be globular clusters belonging to NGC~1291.
A significantly higher percentage of the X-ray sources in the outer regions
of NGC~4697 and NGC~1553 are potentially associated with a globular
cluster than was found for NGC~1291.
However, the studies of NGC~4697 and
NGC~1553 benefited from globular cluster lists derived from ground-based
or {\it Hubble Space Telescope} data,
so it was possible to identify fainter candidate globular
clusters. We were unable to find a globular cluster list for NGC~1291
in the literature, and it is likely that several more of the X-ray sources
detected in the outer regions of NGC~1291 also belong to globular clusters.

\subsection{Diffuse Emission} \label{ssec:diffuse}

Since the soft band X-ray emission is roughly circularly symmetric, we extracted
a radial surface brightness profile in circularly-concentric annular rings
centered on the center of NGC~1291, and excluded all point sources detected
at the $>3\sigma$ level. The soft band emission was detectable out to a
radius of 120$^{\prime\prime}$, after which the emission flattened to a
constant value. We chose an annular region
140$^{\prime\prime}-170^{\prime\prime}$ in extent as our background
region, being sure to omit the part of this region that was not covered by
both observations. The raw and the background-subtracted surface brightness
profiles for the soft band are shown in Figure~\ref{fig:soft_profile}.

XSPEC simulations show that 0.32 keV gas (which describes the spectrum of
the soft emission; see \S~\ref{ssec:diffuse_spec} below),
emits a negligible amount of flux above 1.5 keV.
Therefore,
we can safely assume that the diffuse hard emission is from unresolved
point sources below the detection threshold of the observation. In fact,
extrapolating the luminosity function (see \S~\ref{ssec:sources}) of the
resolved sources down to $10^{35}$ ergs s$^{-1}$ predicted a value for the
luminosity for the unresolved sources that was within 4\% of the
\centerline{\null}
\vskip3.35truein
\includegraphics{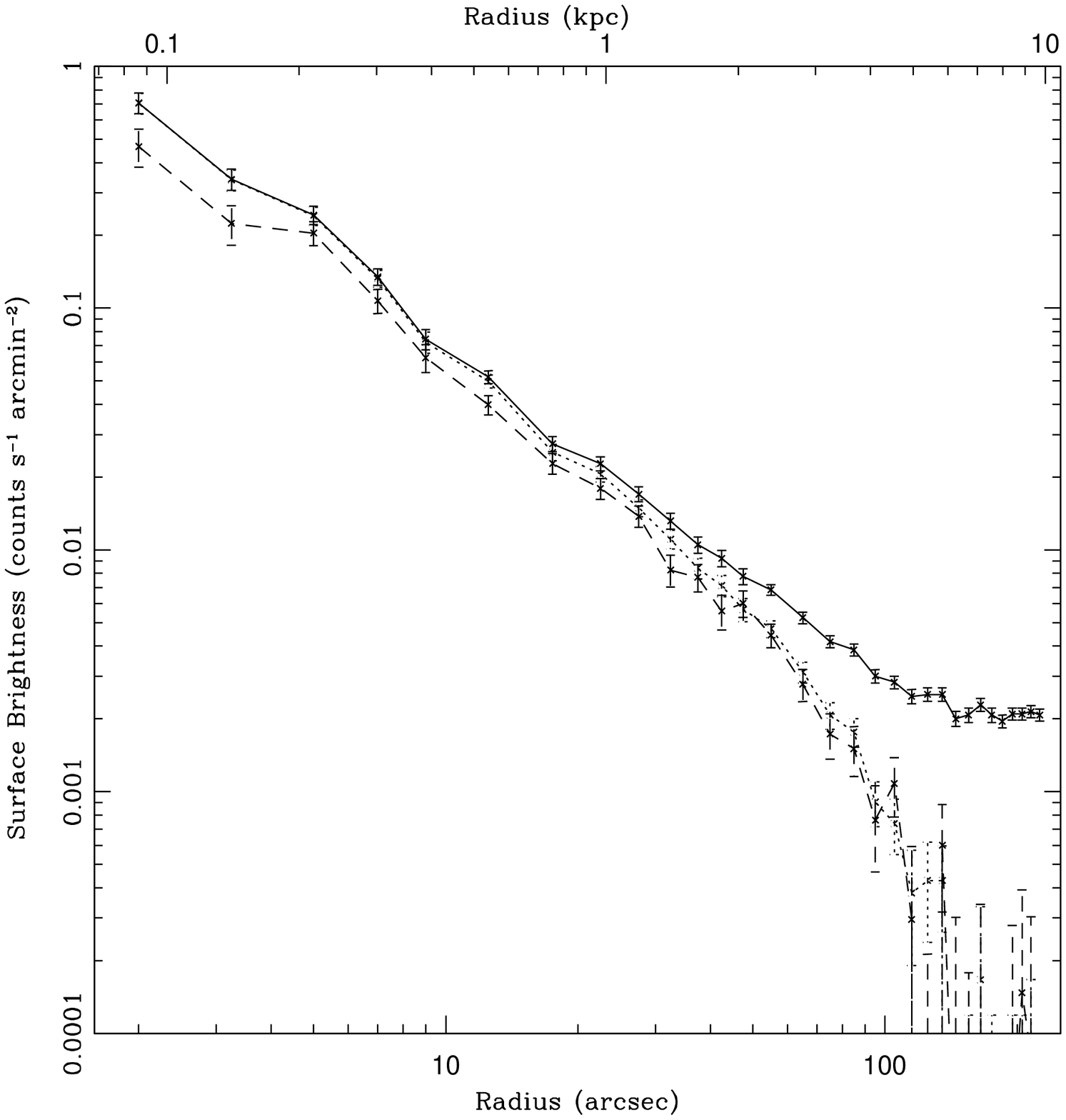}
\figcaption{\small
Radial surface brightness of the diffuse emission
in the 0.3--1.0 keV band, showing the raw profile ({\it solid line}),
background-subtracted profile ({\it dotted line}), and unresolved
source-corrected profile ({\it dashed line}).
\label{fig:soft_profile}}
measured value for the diffuse hard emission. The diffuse
emission in the soft band, however, is a combination of hot gas and unresolved
sources. We can use the observed radial profile of the 1.5--6 keV emission
to correct the 0.3--1.0 keV radial profile for stellar contamination. We
assume that the ratio of 0.3--1.0 keV/1.5--6.0 keV count flux is the same
for the unresolved point sources as it is for the resolved point sources,
and multiply the hard radial profile by this ratio
\centerline{\null}
\vskip3.35truein
\includegraphics{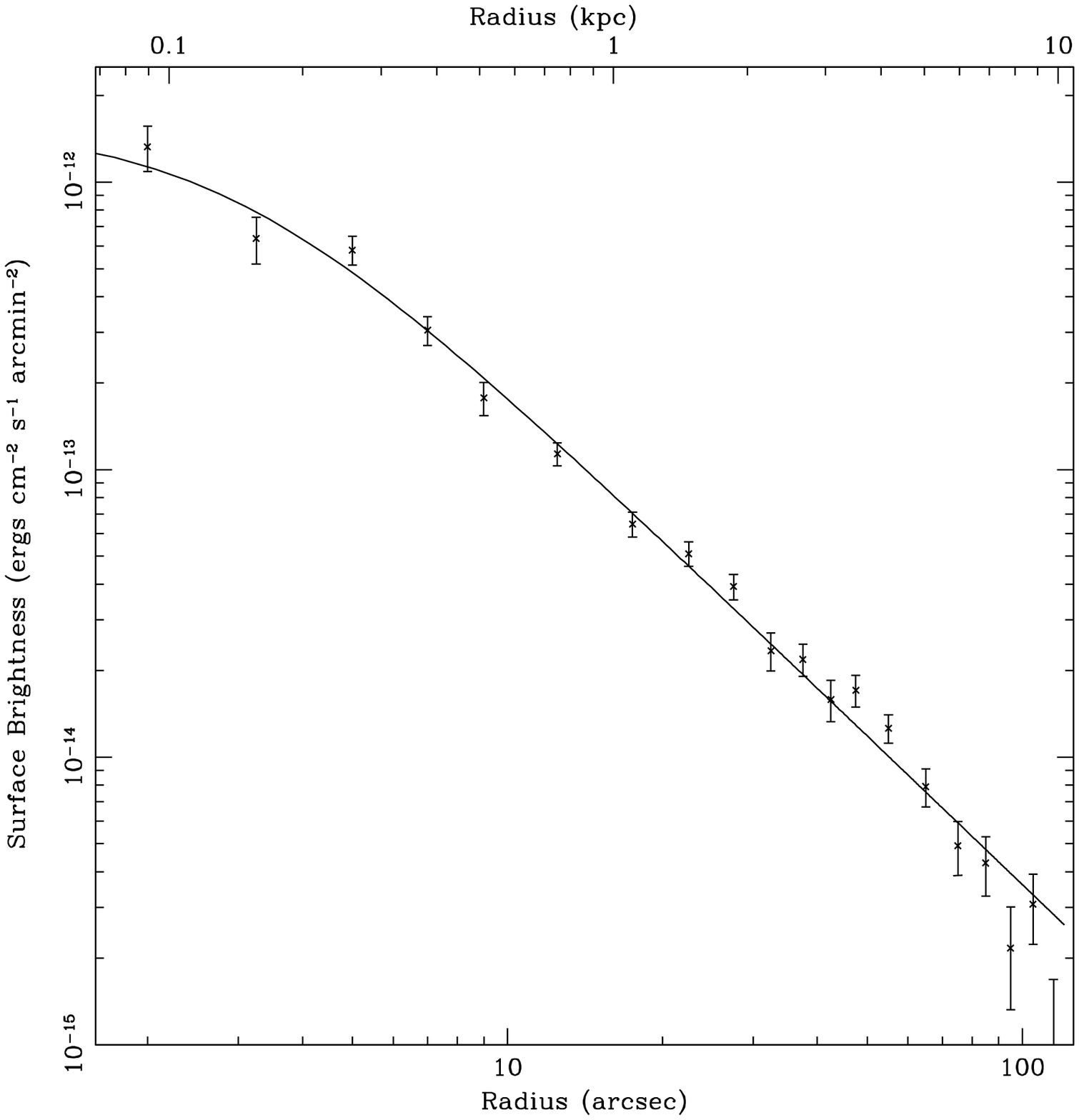}
\figcaption{\small
Radial surface brightness of gaseous emission in NGC~1291,
with best-fit $\beta$ profile $\beta=0.45$ and $r_{\rm core}=2\farcs9$
(125 pc).
\label{fig:xsurf}}
before subtracting it
from the soft radial profile to yield the gas-only radial profile. This profile
is also shown in Figure~\ref{fig:soft_profile}. We fit a $\beta$-profile to the
gaseous surface brightness distribution and found best-fit values of
$\beta=0.45^{+0.02}_{-0.01}$ and $r_{\rm core}=2\farcs9^{+1.1}_{-0.9}$
($0.125^{+0.043}_{-0.039}$ kpc) (Figure~\ref{fig:xsurf}). A value of
$\beta=0.45$ is typical for elliptical galaxies
(Forman et al.\ 1985; Trinchieri et al.\ 1986). Previous
studies of elliptical galaxies have found much larger core radii than was found
for NGC~1291. However, all these galaxies were quite large, luminous, and
rich in hot gas. A recent {\it ROSAT} HRI analysis of several smaller,
X-ray faint galaxies, found core radii similar to the value found here
(Brown \& Bregman 2001).
\centerline{\null}
\vskip3.35truein
\includegraphics{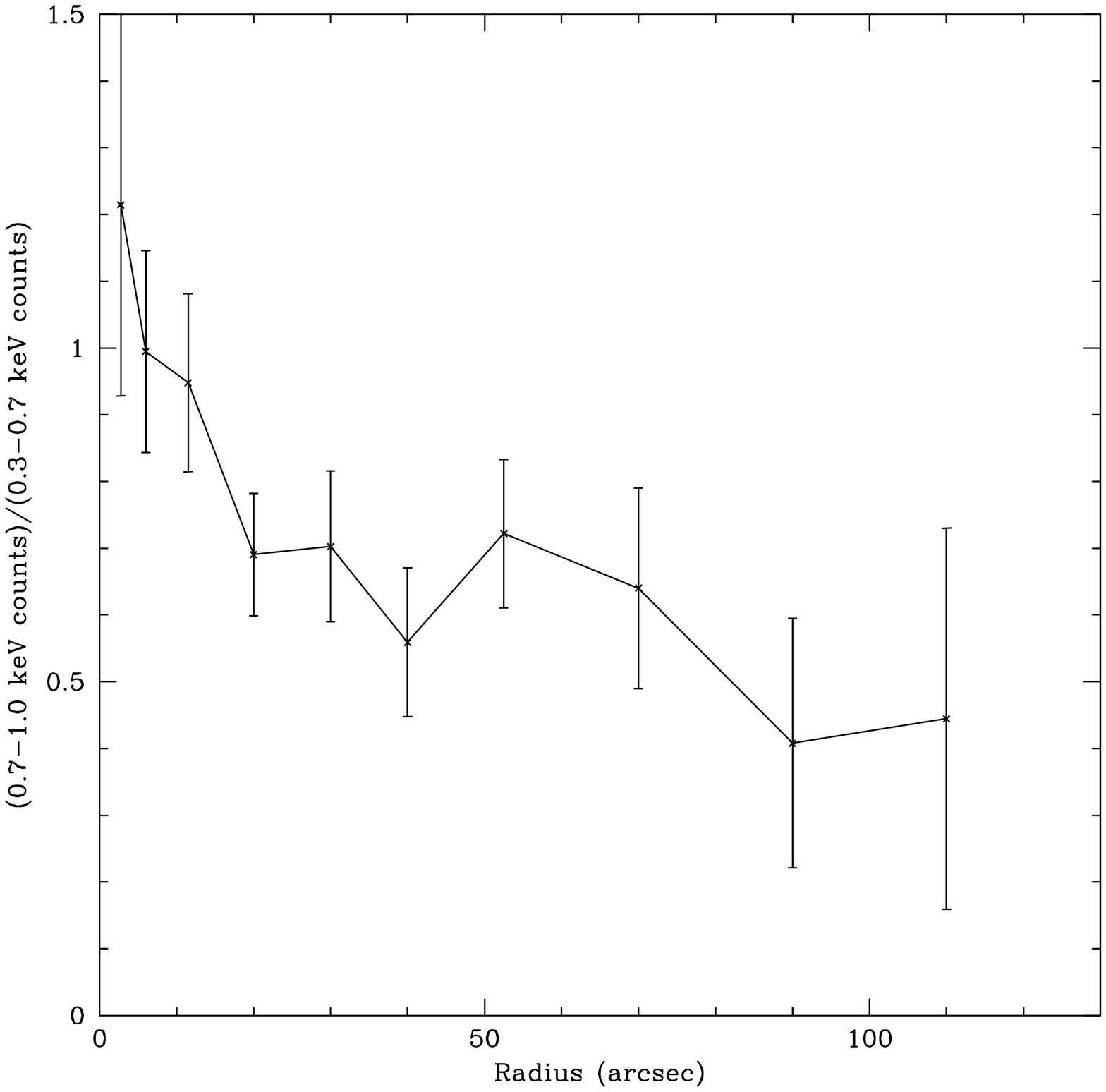}
\figcaption{\small
Ratio of 0.3--0.7 keV counts to 0.7--1.0 keV counts as a function of radius
(1$\sigma$ uncertainties).
The softening of the spectrum with radius indicates a decrease in temperature
or a decrease in metallicity, or both.
\label{fig:ratio}}

\begin{table*}[b]
\caption{X-ray Spectral Fits \label{tab:spectral_fits}}
\begin{center}
\begin{tabular}{llcccl}
\tableline
\tableline
Component &Model& $N_H$ & $kT$ or $\Gamma$&Abundance&$\chi_{\nu}^2$/dof\cr
& & ($10^{20}$ cm$^{-2}$) &(keV)&(relative to solar)& \cr
\tableline
Sources & bremss & (1.60) & $7.5^{+1.6}_{-1.2}$ && 1.17/112\cr
Sources & bremss & $<0.64$& $8.9^{+2.1}_{-1.6}$ && 1.11/111\cr
Sources & power  & (1.60) & $1.56\pm0.05$ && 1.06/112\cr
Sources & power  & $<3.82$ &$1.56\pm0.09$ && 1.07/111\cr
&&&&&\cr
Gaseous (global) & MEKAL  & (1.60) &$0.33^{+0.05}_{-0.03}$&$0.04\pm0.02$&1.08/17
0\cr
Gaseous (global) & MEKAL  & $<6.02$ &$0.35^{+0.06}_{-0.05}$&$0.04^{+0.04}_{-0.02
}$&1.09/169\cr
Gaseous ($1\farcs5-15^{\prime\prime})$ & MEKAL & (1.60) & $0.35^{+0.05}_{-0.03}$
& $0.10^{+0.03}_{-0.02}$&1.65/31\cr
Gaseous ($15^{\prime\prime}-100^{\prime\prime})$ & MEKAL & (1.60) & $0.30^{+0.04
}_{-0.03}$ & $0.04^{+0.04}_{-0.02}$ & 1.14/155\cr
&&&&&\cr
Central source\tablenotemark{a} & power  & $203.3^{+74.1}_{-69.3}$ &
$1.93^{+0.37}_{-0.28}$ && 0.68/48\cr
Central source\tablenotemark{a} & bremss & $159.4^{+49.9}_{-48.3}$ &
$8.3^{+8.3}_{-3.4}$ && 0.68/48\cr
Central source\tablenotemark{b} & MEKAL & (1.60) & $0.75^{+0.12}_{-0.11}$ &
$0.09\pm0.04$ & 0.76/70\cr
Central source\tablenotemark{b} & blackbody  & (1.60) & $0.21\pm0.03$ && 0.86/70
 \cr

\tableline
\end{tabular}
\end{center}
\tablenotetext{a}{1.5--10 keV only, with normalizations free for OB1 and OB2.}
\tablenotetext{b}{0.5--10 keV; soft component parameters, after fixing
the parameters of the power law component derived in the 1.5--10 keV fit.}
\end{table*}
We extracted the surface brightness profile of the unresolved emission in
the 1.5--6 keV band in elliptically-concentric rings that followed the
optical light of the galaxy, and found that the diffuse hard X-ray emission
was distributed like the optical light (de Vaucouleurs 1975). This is
consistent with the idea that the diffuse hard X-ray emission emanates
from LMXBs with luminosities below the detection threshold.
We also determined the ratio of counts in the 0.7--1.0 keV band to the
counts in the 0.3--0.7 keV band (Figure~\ref{fig:ratio}).
The ratio decreases steeply with radius out to 15$^{\prime\prime}$, and
declines more slowly at larger radii.
This decline represents either a decline in temperature or a decline in
metallicity, and will be investigated further in \S~\ref{ssec:diffuse_spec}.

\section{Spectral Analysis} \label{sec:spectral}

Previous {\it Chandra} studies have found that the calibration of the ACIS-S
chip is currently suspect below 0.7 keV (Markevitch et al.\ 2000;
SIB01), so we initially fit all spectra only in the 0.7--10 keV
energy range. However, it was found that the best-fit 0.7--10 keV spectrum
of the point sources provided an adequate fit to
the spectrum even when channels down to an energy of 0.5 keV were added.
Therefore, we include channels down to 0.5 keV in all subsequent fits.
For the diffuse emission, we limited the spectral
range to 0.5--6.0 keV to eliminate high background rates at high energies.
Background was once again taken from an annular region
140$^{\prime\prime}-170^{\prime\prime}$
in extent. The spectra were regrouped so that each channel contained at
least 25 counts. Spectra were extracted from OB1 and OB2 separately and fit
within XSPEC 11.0 separately. (Redistribution matrix files (RMF) and
ancillary response files (ARF) files were created using the
{\sc calcrmf} and {\sc calcarf} routines of Alexey Vikhlinin
for extended sources
and with the {\sc ciao} task {\sc psextract} for the central point source.
All errors given are 90\% confidence levels for one interesting parameter
($\Delta \chi^2 = 2.71$). A summary of the spectral fits is given in
Table~\ref{tab:spectral_fits}.

\subsection{Summed Resolved Point Sources In the Bulge}
\label{ssec:sources_spec}

In order to avoid excessive contamination from unrelated foreground/background
X-ray sources and from disk/spiral arm population X-ray sources of NGC~1291,
we analyze only sources projected within the bulge of the galaxy.
We also excluded the central source that is most likely an AGN.
As noted above (\S~\ref{ssec:sources}), we would
expect only $\sim$3 serendipitous foreground/background sources above
our detection threshold within
the bulge.
A simple bremsstrahlung model with a temperature of
$kT=7.5^{+1.6}_{-1.2}$ keV and the absorption fixed at the Galactic value
($1.60 \times 10^{20}$ cm$^{-2}$) fit the data adequately
($\chi^2_{\nu} = 1.17$ for 112 degrees of freedom). A power law model provided
a somewhat better fit ($\chi^2_{\nu} = 1.06$ for 112 degrees of freedom) with
$\Gamma=1.56\pm 0.05$. A disk blackbody + blackbody model that has been
employed to fit the spectra of Galactic LMXBs as well as the summed emission
from LMXBs in M31 (Mitsuda et al.\ 1984) provided an equally acceptable fit
($\chi^2_{\nu} = 1.06$), but the best fit parameters ($kT_{in} = 0.37$ keV
and $kT_{BB} = 1.1$ keV) were unlike those of Galactic LMXBs.
Freeing the absorption did not improve the fit. The power law exponent
and the bremsstrahlung temperature are in agreement
with previous {\it Chandra} observations of LMXBs in early-type systems
(SIB01; Blanton et al.\ 2001). No excess soft emission was detected in the
spectra of the LMXBs.
The summed source fluxes from OB1 and OB2 were consistent within the
uncertainties, indicating that the effect of variability of the
summed source spectrum is small.

\subsection{Diffuse Emission} \label{ssec:diffuse_spec}

The color analysis of the diffuse emission within
the bulge indicated the presence
of both stellar and gaseous components in the diffuse emission, so we used
a two component model consisting of a MEKAL component
(Mewe, Gronenschild, \& van den Oord 1985; Kaastra \& Mewe 1993;
Liedahl et al.\ 1995) to represent the gaseous
emission and a power law or thermal bremsstrahlung to represent the unresolved
stellar sources. The hard component parameter ($\Gamma$ or $kT$) was fixed
at the value found for the sum of the resolved sources. The temperature and
metallicity of the MEKAL component were allowed to vary. A good fit
($\chi^2_{\nu} = 1.08$ for 170 degrees of freedom) was obtained with
a temperature of $0.32^{+0.04}_{-0.03}$ keV and metallicity of
$0.06 \pm 0.02$ solar when the hard component was a power law model.
Nearly identical values were found when a bremsstrahlung or disk blackbody +
blackbody model represented the
hard component. The temperature and abundance values were practically
independent of the shape of the hard component. Fixing the exponent of the
power law at $\Gamma=1$ or $\Gamma=2$ led to less than a 5\% change in the
best-fit temperature.

Next, we extracted spectra from annular regions
$1\farcs5-15^{\prime\prime}$ and $15^{\prime\prime}-100^{\prime\prime}$
in extent, and fit them with a similar model as above. From
the inner bin to the outer bin, the temperature
declined very slightly from
$0.34^{+0.05}_{-0.03}$ keV to $0.30^{+0.04}_{-0.03}$ keV.
Interestingly, there was some evidence for decreasing abundance with radius
from $0.13 \pm 0.04$ to $0.05^{+0.04}_{-0.02}$ of the solar value.

\subsection{Central Point Source} \label{ssec:central_spec}

The spectrum from the bright central point source could not be fit well
with any single component model in the 0.5--10 keV energy range. However,
the emission within the $1\farcs5$ extraction radius is likely contaminated
by $\sim$0.34 keV gaseous emission. Therefore, we fit the spectrum of the
central source only over the 1.5--10 keV range where the gaseous emission
is negligible. Now the spectrum is fit equally well by a heavily absorbed
power law ($\Gamma=1.93^{+0.37}_{-0.28}$ and $N_H=2.03^{+0.74}_{-0.69} \times
10^{22}$ cm$^{-2}$) or thermal bremsstrahlung ($kT=8.3^{+8.3}_{-3.4}$ keV and
$N_H=1.59^{+0.50}_{-0.48} \times 10^{22}$ cm$^{-2}$). The power law
exponent/bremsstrahlung temperature
and the absorption were linked for OB1 and OB2, and allowing them to vary
separately did not significantly improve the fit. These values are quite unlike
those of the summed non-central point source spectrum in NGC~1291, which was fit
well by a thermal model of similar temperature but with only Galactic
absorption. The 2--10 keV luminosity of the central source decreased by a
factor of two between OB1 and OB2, from $2.9 \times 10^{39}$ ergs s$^{-1}$ to
$1.4 \times 10^{39}$ ergs s$^{-1}$ (for the power law model).
The high variability,
a power law index of 1.7--2.0, and heavy ($10^{22}$ cm$^{-2}$) absorption
argues that the source is an obscured low luminosity AGN.

\vskip2.55truein
\includegraphics{f8.eps}
\figcaption{\small
Best-fit power law spectrum of the central point source from
OB1 and OB2 over the 1.5--10 keV range. Energy channels below 1.5 keV
have been added without re-fitting the spectrum to illustrate the excess
soft emission. Note that although the power law component differs by a factor
of two between OB1 and OB2, the excess soft component flux is the same between
the two observations.
\label{fig:central}}
After freezing the best-fit parameters, we included channels down to
0.5 keV and noticed excess emission in the 0.5--1.5 keV range
(Figure~\ref{fig:central}). A total of 285 counts were found in excess of the
absorbed power law model. Below 1.0 keV the count rates of OB1 and OB2 were
the same, indicating a different source of the soft emission than what is
producing the variable hard emission. The number of counts below 1.5 keV
that is expected within the $1\farcs5$ extraction region from the hot gas
(by extrapolating the surface brightness profile of the hot gas to the very
center) is only $\sim50$ counts. Thus, most of the excess counts must
come from a second, soft component of the central source. After fixing the
absorption and power law exponent, we added a MEKAL component with a
temperature of 0.34 keV to model the hot ISM gas expected in the
$1\farcs5$ extraction region, and fixed the normalization so that this
component yielded 50 counts. We then added another MEKAL model and allowed
the temperature and metallicity to vary. This component was absorbed
by a column density equal to the Galactic column density in that direction.
The best-fit MEKAL temperature was $kT=0.75^{+0.12}_{-0.11}$ keV
and the best-fit metallicity was $0.09\pm 0.04$ solar
($\chi^2_{\nu} = 0.76$ for 69 degrees of freedom). Substituting a
blackbody model for the second MEKAL model yielded a somewhat worse but still
acceptable fit with $kT_{BB}=0.21\pm0.03$ keV. The luminosity of
the soft component was $1.5 \times 10^{38}$ ergs s$^{-1}$, or 5\%--10\% of
the hard component luminosity.

\section{Discussion} \label{sec:discussion}

\subsection{A Comparison of the Stellar $L_X/L_B$ Ratio of NGC~1291,
NGC~4697, and NGC~1553} \label{ssec:XLF}

Previous work has indicated that the stellar X-ray--to--optical luminosity
ratio is not the same in all galaxies. In a survey of 61 early-type
galaxies observed by {\it ROSAT}, Irwin \& Sarazin (1998b) found that eight
galaxies had extremely low $L_X/L_B$ values, over a factor of three less
than the $L_X/L_B$ value of the bulge of M31 and other X-ray faint early-type
galaxies whose X-ray emission we now know to be composed primarily of
of LMXBs with only a small amount of gas such as NGC~4697
(Primini et al.\ 1993; Shirey et al.\ 2001; SIB01).
If all the gas were removed from NGC~4697, for example,
its $L_X/L_B$ ratio would decrease by only a factor of $\sim1.4$ in the
{\it ROSAT} band. Since eight galaxies in the Irwin \& Sarazin (1998b) sample
had $L_X/L_B$ values at least a factor of three below that of M31 or NGC~4697,
this implied that the production of LMXBs in early-type systems is
different from galaxy to galaxy, and that the X-ray luminosity from stellar
sources does not scale linearly with the optical luminosity. The spatial
resolution of {\it ROSAT} was too coarse to test this hypothesis directly,
since individual LMXBs could not be resolved. It was not known if the
discrepancy in the LMXB population resulted from a lack of LMXBs at all
luminosities or just at the high end of the luminosity distribution.

{\it Chandra's} superb spatial resolution allows us to distinguish between
the two scenarios.
The total (resolved plus unresolved) stellar $L_X/L_B$ value for NGC~1291
is estimated to be $5.3 \times 10^{29}$ ergs s$^{-1}$ $L_{\odot}^{-1}$
(of which 71\% is resolved in our observation).
In the 0.5--2.0 keV ({\it ROSAT}) band, this stellar $L_X/L_B$ value is
$1.5 \times 10^{29}$ ergs s$^{-1}$ $L_{\odot}^{-1}$.
Using the source list as well as the estimated emission from unresolved
point sources from the {\it Chandra} observation of NGC~4697
(SIB01), we found the total stellar $L_X/L_B$ value for
NGC~4697 within one effective (half-optical light) aperture was
$7.5 \times 10^{29}$ ergs s$^{-1}$ $L_{\odot}^{-1}$ (it is easiest to deal only
with sources within the effective aperture in large galaxies like NGC~4697,
since the number of contaminating serendipitous sources is much smaller
within this angular region than from over the entire optical extent of the
galaxy). A similar exercise within two effective radii for NGC~1553
(Blanton et al.\ 2001) yielded a
value of $7.2 \times 10^{29}$ ergs s$^{-1}$ $L_{\odot}^{-1}$. In all three
galaxies, the central source was omitted. From the luminosity distribution
function of NGC~1291 (Figure~\ref{fig:xlf}) it is apparent that this difference
can be attributed to a lack of bright sources in NGC~1291. In fact there are
no sources in NGC~1291 brighter than $3 \times 10^{38}$ ergs s$^{-1}$.
Direct comparison with NGC~4697
and NGC~1553 indicates that the lack of sources above
$3 \times 10^{38}$ erg s$^{-1}$ in NGC~1291 is probably not the result of low
number statistics. Within the inner effective apertures
of NGC~4697 and NGC~1553, there are 10 sources each with luminosities
greater than $3 \times 10^{38}$ erg s$^{-1}$ (neglecting central sources that
might be AGN).
Given that the ratio of
optical luminosities within the effective aperture of NGC~4697 and NGC~1553 to
the optical luminosity of the entire bulge of NGC~1291 is 1.17 and 2.34,
respectively, we would expect 3.7--7.3 sources in NGC~1291 above this
luminosity. If we include all sources detected in the field, the number of
sources above this luminosity grows to 19 and 35 for NGC~4697 and NGC~1553,
respectively, again predicting $\sim7$ sources in the bulge of NGC~1291.
In NGC~4697, NGC~1553, and NGC~1399
(R. Mushotzky \& L. Angelini, private communication) there
was a break or ``knee" in the luminosity distribution function around a
luminosity of $3 \times 10^{38}$ erg s$^{-1}$. This luminosity is also
approximately the Eddington luminosity of a 1.4 M$_{\odot}$ neutron star.
SIB00 suggest that this break represents a division between
LMXBs with black hole and neutron star accreters.
If this interpretation is correct,
it would mean that NGC~1291 does not have any black hole
binaries accreting near the Eddington limit. It is worth noting that the
inner $5^{\prime}$ of the bulge of the spiral galaxy M31 also does not
contain any X-ray source more luminous
than $3 \times 10^{38}$ ergs s$^{-1}$ (Primini et al.\ 1993; after converting
their 0.2--4.0 keV luminosity to 0.3--10 keV luminosity).

Another illustrative way of presenting the luminosity distribution function
is shown in Figure~\ref{fig:cumul}{\it a}, where we have plotted the cumulative
$L_X/L_B$ value of sources in the galaxy as a function of increasing $L_X$.
Since we have
normalized the X-ray luminosity of the sources by the optical luminosity of the
galaxy we can now compare the curve to those of other galaxies.
Figure~\ref{fig:cumul}{\it a} also shows the same quantity for the inner
effective aperture of NGC~4697 and within two effective radii for NGC~1553.
Note that the y-intercept of each curve represents the {\it unresolved}
source $L_X/L_B$ value, a quantity set by the limiting luminosity
of the observation (i.e., the curves would go to zero if every last faint LMXB
could be detected). Since the determination of $L_B$ is crucial in this
comparison, we did not attempt to gauge $L_B$ from the bulge assuming a
specific bulge-disk decomposition for NGC~1291. Instead we only included X-ray
sources within a circular aperture of $131\farcs0$, for which the
integrated $B$ magnitude within this aperture was 10.13 (using photometry
values listed in the compilation of Prugniel \& Heraudeau 1998). The
Galactic reddening maps of Schlegel, Finkbeiner, \& Davis (1998) indicated
an absorption of $A_B=0.056$ magnitude. Correcting for foreground reddening
and assuming a distance of 8.9 Mpc, we determined an optical luminosity
of $1.08 \times 10^{10}$ $L_{\odot}$. Schlegel et al.\ (1998) claim a mean
uncertainty of 16\% of $A_B=0.056$, or 0.009 magnitude. This amounts to
only a 1\% uncertainty in $L_B$ that results from uncertainties associated
with foreground extinction corrections.

Two striking features stand out in Figure~\ref{fig:cumul}{\it a}.
The total cumulative
$L_X/L_B$ value of NGC~1291 is a factor of 1.4 less than NGC~4697 and NGC~1553.
As mentioned above, this is the result of a lack of very luminous sources in
\centerline{\null}
\vskip6.00truein
\includegraphics{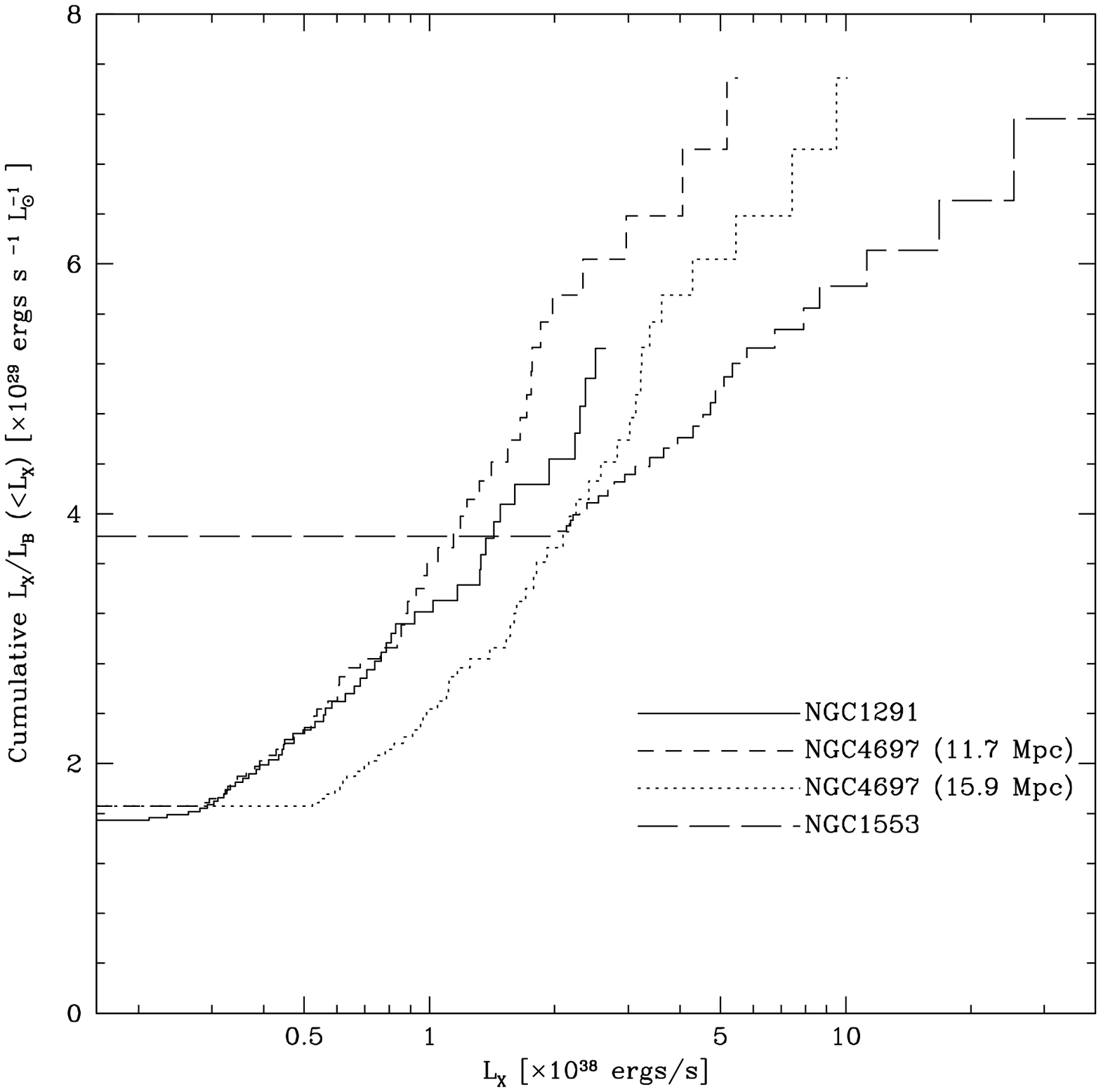}
\includegraphics{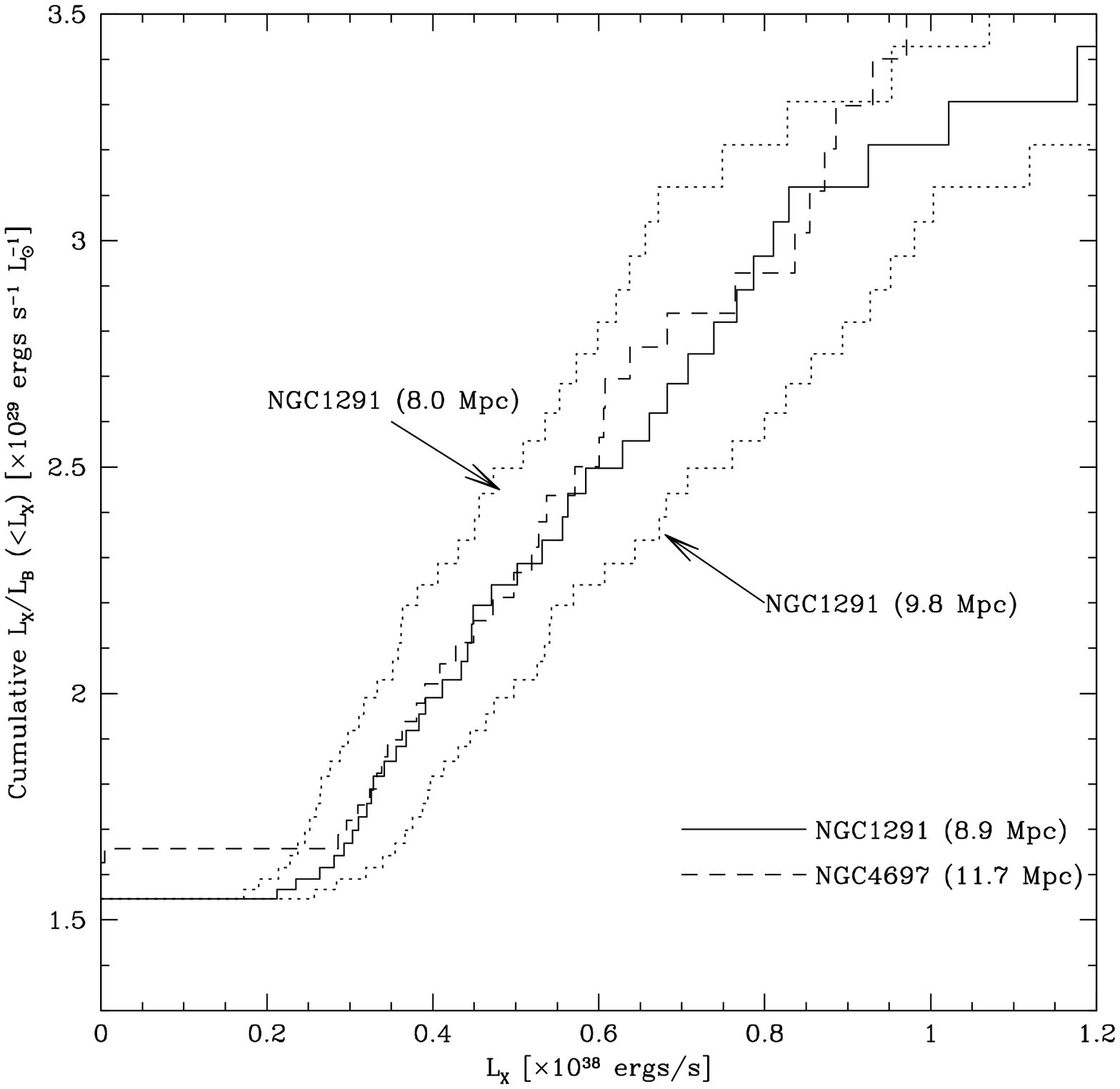}
\figcaption{\small
({\it a}) Cumulative $L_X/L_B$ values as a function of increasing $L_X$ for
LMXBs in NGC~1291, NGC~4697 (for distances of 15.9 Mpc and 11.7 Mpc),
and NGC~1553. Note the disparity in the
number of high luminosity LMXBs among the three galaxies.
({\it b}) A closer view of the lower left corner of
Figure~\protect\ref{fig:cumul}{\it a}, showing the agreement in the
cumulative $L_X/L_B$ value of LMXBs in NGC~1291 and NGC~4697 (when a distance
of 11.7 Mpc is chosen). Changing the
distance of NGC~1291 by $\pm10\%$ destroys this agreement, and suggests
that this agreement might be used as a relative distance indicator.
\label{fig:cumul}}
NGC~1291. This is surprising considering the optical characteristics
of early-type systems are very similar. However, for some reason
NGC~1291 does not have any of the brightest (probable) black hole binaries
that other galaxies have. All three galaxies show a substantial spread
in the number and luminosity of sources above $3 \times 10^{38}$ ergs s$^{-1}$,
the break luminosity discussed above. Also, below $10^{38}$ ergs s$^{-1}$
the slope of the cumulative $L_X/L_B$ values of NGC~1291 and NGC~4697 are
quite similar (NGC~1553 is too distant and the observation time was
too short to test this with this galaxy too). In fact, if the distance
to NGC~4697 is changed from 15.9 Mpc (the distance assumed by SIB01, which
is the recessional velocity distance determined using a value of
$H_0=50$ km s$^{-1}$ Mpc$^{-1}$; Faber et al. 1989) to 11.7 Mpc (the distance
implied by the surface brightness fluctuation method; Tonry et al.\ 2001),
the agreement in the cumulative $L_X/L_B$ profiles of NGC~1291 and NGC~4697
at low luminosities
is remarkable. This indicates that the number of faint LMXBs agrees quite well
between the two galaxies, if the smaller distance for NGC~4697 is adopted.

The discrepancy in the number of luminous sources in NGC~1291 can be
reconciled with those of NGC~4697 and NGC~1553 if we assume a larger
distance to NGC~1291 than 8.9 Mpc. If we increase the
distance by 40\% to 12.5 Mpc, we find nine sources above
$3 \times 10^{38}$ erg s$^{-1}$. However, this creates another problem.
Although the total cumulative $L_X/L_B$ value is independent of distance,
the curve
shifts to the right when the distance is increased. When the distance to
NGC~1291 is changed from 8.9 Mpc, the agreement in the cumulative $L_X/L_B$
value below $10^{38}$ ergs s$^{-1}$ is destroyed.
Regardless of the chosen distance, the data
conclusively show that the production of LMXBs in NGC~1291 differs from
that in NGC~4697. If the distance is set to match
the low end of the luminosity distribution function, too few bright
sources are produced in NGC~1291. If the distance is set to match the
high end, too few fainter ($< 10^{38}$ erg s$^{-1}$) sources are
produced in NGC~1291.
Moreover, the shape of the upper end of the luminosity function of NGC~1291
disagrees with that of NGC~4697
(\S~\ref{ssec:sources}), independent of the distance.

Assuming the distances of 8.9 Mpc and 11.7 Mpc to NGC~1291 and NGC~4697,
respectively, are correct, the excellent agreement
in the cumulative $L_X/L_B$ value below
$10^{38}$ ergs s$^{-1}$ between NGC~1291 and NGC~4697
(Figure~\ref{fig:cumul}{\it a})
suggests that this feature can be used to estimate the distance to
nearby galaxies. If the cumulative $L_X/L_B$ value below
$10^{38}$ ergs s$^{-1}$ is a constant from galaxy to galaxy, the point where
the cumulative $L_X/L_B$ value begins to diverge might be used as a standard
candle. Using X-ray binaries as a standard candle is not a new concept.
Margon \& Ostriker (1973) suggested that the luminosity of X-ray sources
in galaxies could be limited by the Eddington luminosity based on
observations of X-ray sources in the Local Group. More recently,
SIB00 suggested that the break in the luminosity
distribution function at $3 \times 10^{38}$ ergs s$^{-1}$ might be used
as a distance indicator too. The accuracy of this method, however, would
always be limited by the relatively small number of sources above the
break.
The statistics on sources below the break are generally much better,
and are further improved by the summing up the luminosities.
What is required observationally is to resolve the
luminosity function down to $\la$$10^{38}$ ergs s$^{-1}$, and to
determine accurately the total amount of X-ray emission from unresolved
sources.
This method could be used in conjunction with the break in the luminosity
function method of SIB00 for greater accuracy. Figure~\ref{fig:cumul}{\it b}
zooms in on the lower left corner of Figure~\ref{fig:cumul}{\it a}, illustrating
how sensitive the cumulative $L_X/L_B$--$L_X$ relation is to distance.
Changing the assumed distance of NGC~1291 by $\pm10\%$ leads to an easily
measurable difference in the cumulative $L_X/L_B$ values of NGC~1291 and
NGC~4697.
Of course claims of the validity of this method are premature
considering that only two galaxies have been tested thus far. Many more
galaxies will need to be compared to NGC~1291 and NGC~4697 to confirm
this method. However,
should this method be valid, distances to nearby galaxies can
be determined to better than 10\% accuracy, and would be one of the very
few non-optical methods for determining distances to galaxies.
We point out that the cumulative $L_X/L_B$ method is a relative distance
estimator, while the method using the break in the luminosity function
is an absolute distance estimator.
Admittedly, the distance to NGC~1291 is not known with great
accuracy, but if this distance estimator method works, we can determine
accurate relative distances to these galaxies. In future work, we will study
galaxies for which multiple distance estimators already exist and agree
with one another and apply our method to them. Also useful will be galaxies
that are known to be at the same distance (owing to their group membership).

Recently, White, Kulkarni, \& Sarazin (2002; in preparation) have found
a correlation between the total LMXB $L_X/L_B$ value
and the specific frequency of globular clusters in a galaxy.
In addition, Angelini, Loewenstein, \& Mushotzky (2001)
found that 70\% of the X-ray point sources found in a portion of NGC~1399
for which {\it Hubble} data existed were coincident with globular clusters
of NGC~1399. Both these results would seem to be at odds with the idea that
the cumulative $L_X/L_B$ profile below $10^{38}$ ergs s$^{-1}$ is uniform from
galaxy to galaxy. One might expect $L_X$ to scale with the globular
cluster specific frequency in addition to the optical luminosity $L_B$
rather than $L_B$ alone, rendering the distance
estimator method useless in its current form. However, it is possible that both
theories are correct. It is known that the specific frequency of globular
clusters decreases with decreasing radius in galaxies (Forbes et al. 1996),
indicating that the spatial distribution of the globular clusters is more
extended than the optical light.
For example, the specific frequency of globular clusters in NGC~3115 decreases
from $3.1 \pm 0.5$ to $1.3 \pm 0.1$ in moving from the outer to the inner
regions of the galaxy (Kundu \& Whitmore 1998). Therefore, globular cluster
X-ray sources might be relatively rare in the inner regions of elliptical
galaxies (or bulges), while comprising a larger percentage of the
X-ray source population at larger radii (20\% in the case of NGC~4697; SIB01).
If this is the case, galaxies with high globular cluster specific frequencies
would have higher $L_X/L_B$ values than galaxies with low globular cluster
specific frequencies because of a higher number of globular cluster X-ray
sources at large radii. At smaller radii (such as within the bulge or one
effective radius of a galaxy) where non-globular cluster X-ray sources
dominate, the $L_X/L_B$ value might be a constant from galaxy to
galaxy, allowing the distance indicator method to be utilized.
A search for globular clusters in the inner regions of elliptical galaxies
with {\it Hubble Space telescope} would be useful in definitively determining
the significance
of globular cluster X-ray sources to the total LMXB population and
resolving this issue.

\subsection{The Hot Gas within NGC~1291} \label{ssec:hot_gas}

The ability to separate the stellar emission from gaseous emission has only
become possible since the launch of {\it Chandra}. Whereas in X-ray bright
galaxies the contribution from LMXBs is minimal, such a separation
becomes important as the gaseous component becomes less dominant. For example,
the {\it ROSAT} PSPC spectrum of NGC~1291 could be fit equally well with
a two-component Raymond-Smith + bremsstrahlung model with
$kT_{\rm RS}=0.15$ keV,
$Z = Z_{\odot}$ and $kT_{\rm brem}=1.07$ (Bregman et al.\ 1995) as it
could with just a Raymond-Smith model with $kT_{RS}=0.55$ keV and
$Z$ = 0.01 solar (Read, Ponman, \& Strickland 1997). In both cases, the
inability of {\it ROSAT} to resolve out the binaries and especially the
luminous, highly absorbed AGN led to an erroneous measurement of the
temperature of the diffuse hot gas. Inaccuracies in the temperature and flux
of the hot gas subsequently adversely affects the determination of quantities
such as the $L_{X,gas}-L_B$ relation and the $kT_{gas}-\sigma$ relation,
both of which have been used to characterize the history and evolution
of the hot gas within early-type systems.

Given the spectral and spatial results above, we conclude that 33\% (41\%) of
the counts and 14\% (27\%) of the 0.3--10 keV (0.3--2 keV) luminosity emanates
from a truly diffuse, gaseous component.
The 0.3--10 keV X-ray luminosity of the gas is
$1.4 \times 10^{39}$ ergs s$^{-1}$, leading to a gaseous $L_X/L_B$ value
of $1.3 \times 10^{29}$ ergs s$^{-1}$ $L_{\odot}^{-1}$. Such a value is quite
low compared to X-ray bright elliptical galaxies, but similar to the values
found for X-ray faint ellipticals and the bulge of M31 (SIB01;
Shirey et al.\ 2001). Given the velocity dispersion of the bulge of NGC~1291
of 162 km s$^{-1}$ (Dalle Ore et al.\ 1991), the derived temperature of
the gas of 0.32 keV falls very close to the best-fit $kT_{gas}-\sigma$
relation of Davis \& White (1996), but considerably above the best-fit
relation of Brown \& Bregman (1998). The measured temperature is well
above the temperature predicted from the velocity dispersion of 0.17 keV,
if the thermal energy of the gas were equal to the kinetic energy of the
random motions of the stars.

The beta-model for the surface brightness profile of the gas can be
inverted to obtain the
density profile, $n(r)=n_0[1+ (r/r_{\rm core})^2]^{-3\beta/2}$.
Using the best-fit values of the gaseous surface brightness profile of
$\beta=0.45$ and $r_{\rm core}=2\farcs9$, we derive
a central electron density of $n_0= 0.28$ cm$^{-3}$. This value is quite high
when compared to values obtained for early-type systems with previous
instruments (e.g., Rangarajan et al.\ 1995).
However, the $0\farcs5$ spatial resolution of {\it Chandra} coupled with
the nearness of NGC~1291 allows us to probe the electron density profile on the
scale of $\sim$50 pc (note that at the distance of the Virgo cluster the
core radius would be only $1\farcs3$, far below the resolution limit of
previous X-ray telescopes). The implied mass of the gas
(assuming a volume filling factor of one) within a radius
of 120$^{\prime\prime}$ is $5.8 \times 10^7$ M$_{\odot}$. This value is
substantially smaller than the derived gas masses of big bright elliptical
galaxies, which tend to be as high as $10^{11}$ M$_{\odot}$. The cooling time
of the gas at the center of the galaxy is $1.2 \times 10^7$ years, and
remains less than a Hubble time out to the extent to which the gaseous
emission can be detected ($120^{\prime\prime}$).

Given the relative isolation of NGC~1291, it would appear that the
lack of a substantial amount of X-ray emitting gas in the bulge is not
the result of recent interaction with a cluster or another galaxy. This
result is in accordance with the findings of Brown \& Bregman (2000)
that optically less luminous galaxies in lower density environments
are X-ray faint. In smaller galaxies (or bulges) in non-cluster environments,
supernovae-driven winds can clear the system of much of the gas that might
have been pressure-contained by intracluster gas had the galaxy resided
in a cluster environment.

Finally, the metallicity of the hot gas in early-type systems has been a
longstanding controversy. Optical measurements of the metallicity of stars
indicate abundances of solar or greater in elliptical galaxies and the
bulges of spiral galaxies. Yet, many measurements of the metallicity of
the X-ray gas, particularly those determined with {\it ASCA}, have yielded
values of the metallicity of $\la20\%$ for many galaxies. Several
explanations have been set forth to account for this discrepancy.
Buote (1999) suggests that the low measured metallicity of the X-ray gas
with {\it ASCA} is an artifact of fitting the spectra with isothermal
models.
Multi-temperature models, either due to local multiphase gas or a temperature
gradient, would result in higher abundances,
at least in the case of the brightest X-ray elliptical galaxies.
Finoguenov \& Jones (2000) also point out
that the angular scales of the optical and X-ray metallicity are different;
whereas integrated optical metallicities are typically measured within
an arcminute or less for galaxies at the distance of Virgo, X-ray
measurements are typically average over $\sim$$3^{\prime}$ bins. Given that
the optically-determined metallicities often decline significantly with
radius (Carollo, Danziger, \& Buson 1993), the stellar metallicity might
actually be subsolar at large distances for the center of the galaxy.
Furthermore, the {\it Chandra}-determined abundance profile of the
elliptical galaxy NGC~4374 peaked toward the center to a value of
0.6 solar within a 0.5--5 kpc annulus, and declined at larger radii
(Finoguenov \& Jones 2001). Both the multi-component temperature argument
and the mismatch of angular size argument seem to resolve the discrepancy
between the metallicity measurements in the much-studied X-ray bright early-type
galaxies. However, neither explain the low metallicity found here for
NGC~1291. The metallicity from the $1\farcs5-15^{\prime\prime}$ annulus
is only $0.13 \pm 0.04$ solar. An optical measurement of the
metallicity of the bulge found it to have an Mg$_2$ index of 0.24 magnitudes
(Golev \& Prugniel 1998) corresponding to Fe/H = 1.1 solar
(Terlevich et al.\ 1981).
In addition, there appears to be no temperature gradient or evidence for
cooling gas in the center of NGC~1291 to bias the metallicity determination
of the gas. It appears that X-ray faint systems like NGC~1291 (and NGC~4697,
although only a global metallicity was determined; SIB01)
really do have low metallicity gas even at their centers.

\subsection{The Central Source}
\label{ssec:central}

The presence of a bright, central X-ray point source is somewhat surprising
considering there is little evidence in the literature that NGC~1291 hosts
an AGN. While not as luminous as other low luminosity AGN, the spectral
parameters of the central X-ray source are very similar to those found for a
sample of low luminosity AGNs, LINERS, and starbursts observed by {\it ROSAT}
and {\it ASCA} by Ptak et al.\ (1999).
Such spectral parameters are also common
among narrow-line Seyfert 1 galaxies (Reeves \& Turner 2000). It should be
noted that in these systems the X-ray luminosity of the AGN is at least a factor
of ten higher than the AGN in NGC~1291.
What is even more surprising is that the flux of the soft component was
unchanged between OB1 and OB2, while the hard component decreased in
intensity by a factor of two over the same time period. The excellent
spatial resolution of {\it Chandra} confirms that the origin of the soft
component lies within $\sim60$ pc from the central source, and is almost
assuredly associated with the central source.

This kind of temporal behavior in the soft and hard components has also been
seen in the narrow-line Seyfert 1 galaxy
Mrk 766 with {\it XMM-Newton} by Page et al.\ (2001).
These authors find
that the soft X-ray excess, as well as the ultraviolet emission observed
with {\it XMM-Newton's} Optical Monitor varied very little throughout the
observation, whereas the hard continuum component varied significantly.
It is unlikely that the soft component can be reprocessed hard power law
emission, since in this case the fluxes of the hard and soft components
should vary together. It is possible that the soft emission emanates directly
from the hot inner accretion disk (Malkan \& Sargent 1982) and is the
high energy tail of the ultraviolet bump seen in these objects. Alternatively,
the source of the soft X-ray emission may be the Compton up-scattering
of the ultraviolet photons from this disk.

The detection of a low luminosity active nucleus in NGC~1291 also explains
the detection of unresolved 60 micron and 100 micron emission in this galaxy
(Rice et al.\ 1988). Bregman et al.\ (1995) point out that it was
surprising that very little if any CO emission was detected in this galaxy
given the 100 micron flux of 10.7 Jy, based on the observed correlation
of CO with 100 micron emission in other galaxies. It is quite possible that
the 100 micron emission is associated with the AGN and not with dust associated
with star formation in the bulge.

\subsection{Hot vs.\ Cold Gas in Sa Galaxies} \label{ssec:hot_cold}

The ratio of hot to cold gas is a strong function of galaxy morphological
type. The interstellar material in late-type spiral galaxies is
predominantly atomic and molecular with trace amounts of hot gas, while in
elliptical galaxies it is just the opposite, with S0 galaxies having a
roughly equal mixture of the two (e.g., Bregman, Hogg, \& Roberts 1992).
As an Sa galaxy, we would expect NGC~1291 to have more cold gas than hot gas.
This is indeed the case here, where the mass of the hot gas is
$5.8 \times 10^7$ M$_{\odot}$, while the H~{\sc I} gas mass is
$1.3 \times 10^9$ M$_{\odot}$ (van Driel, Rots, \& van Woerden 1988).

The H~{\sc I} distribution in NGC~1291, as is the case with several S0 and
Sa galaxies, is ring-like in nature (van Driel et al.\ 1988). The
central hole in the H~{\sc I} is 7 kpc in radius. An analysis of {\it ROSAT}
PSPC data by Bregman et al.\ (1995) found that the X-ray emission appears
to fill this hole. We confirm this result with our {\it Chandra} observations;
hot, diffuse gas with a temperature of $\sim$0.3 keV is detected out to a
radius of 5 kpc.

Whether the hot gas has actually displaced the cold gas or whether it is
merely the case that cold gas is a disk phenomenon while hot gas is a bulge
phenomenon is still an open question. As Bregman et al.\ (1995) point out,
if the hot and cold gas are thermally coupled, convection and turbulent
mixing can lead to the conversion of hot gas to cold gas and vice versa,
causing the the two phases to be spatially anticorrelated, with the phase
with the higher column density winning out. A signature of this would be
a softening of the spectrum of the hot gas with increasing radius, as
successively more hot gas is cooled by the presence of cold gas. Indeed,
there is a softening of the 0.7--1.0 keV/0.3--0.7 keV ratio with increasing
radius (Figure~\ref{fig:ratio}), although it should be noted that the
softening occurs at a radius in which there is no detectable
H~{\sc i} emission. A more definitive answer might be
obtained with an Sa or S0 galaxy that does not have such a large hole
in its H~{\sc I} distribution. Several galaxies in the H~{\sc I} sample
of van Driel \& van Woerden (1991) show complex H~{\sc I}
morphology in the center of the
galaxy. A high resolution {\it Chandra} observation of the bulges of
these galaxies might shed some light on the interaction (or lack of interaction)
between the hot and cold phases of the interstellar medium. If the two
phases were thermally isolated, the cold gas would absorb soft X-rays from
the hot gas behind it, leading to a hardening of the X-rays in the
direction of the H~{\sc I} knots. Alternatively, if the phases are thermally
coupled, we might expect the the X-ray gas to fill the empty pockets in
the H~{\sc I} distribution, indicating that cold gas has been turned into
the hot phase in these regions.

\section{Conclusions} \label{sec:conclusions}

Analyses of two {\it Chandra} observations of the bulge-dominated spiral
NGC~1291 have revealed the primary X-ray--emitting components of the
galaxy. The bulk of the X-rays (by counts and by luminosity) emanates
from a collection for several dozen LMXBs that exhibit a range of spectral
characteristics. A significantly smaller fraction of the X-ray emission
comes from a 0.32 keV thermal gas with low metallicity. These
observations demonstrate, along with {\it XMM-Newton} observations
of the bulge of M31 (Shirey et al.\ 2001), that the X-ray emission of bulges
of galaxies resemble that of a sub-class of elliptical and S0 galaxies
that are X-ray faint. The rather high temperatures measured for Sa
galaxies with {\it Einstein} (Kim et al.\ 1992) were the result of the
mixture of two spectrally distinct components, namely LMXBs and hot gas.

The luminosity distribution function of LMXBs in the bulge of
NGC~1291 can be compared to that of other early-type galaxies to explore
differences in the LMXB populations from galaxy to galaxy. There are no
LMXBs with luminosities greater than $3 \times 10^{38}$ ergs s$^{-1}$ in
NGC~1291, where about seven were expected by scaling from NGC~4697 or
NGC~1553. This lack of high luminosity LMXBs is the cause of the wide spread
in the stellar $L_X/L_B$ values seen with previous X-ray instruments.
Such a spread suggests that the production of LMXBs with massive
primaries is not uniform from galaxy to galaxy, despite the uniformness
of the present-day stellar population among early-type systems. On the
other hand, below $10^{38}$ ergs s$^{-1}$ there is remarkable agreement
in the LMXB functions of NGC~1291 and NGC~4697. This
agreement is highly sensitive to the assumed distance of NGC~1291.
If this agreement is a generic property of LMXBs in early-type systems,
it can be used as an accurate distance indicator, one of the very few
non-optical methods for determining galactic distances.

The distribution of the hot gas can be described by the standard $\beta$
profile with $r_{\rm core}=2\farcs9$ and $\beta = 0.45$. The central density
of the gas is 0.28 cm$^{-3}$, and the total mass of the gas is
$5.8 \times 10^7$ M$_{\odot}$, far less than the mass of the H~{\sc i} in
the galaxy. There is marginal evidence that the temperature and metallicity
decrease slightly outside of 0.6 kpc. The temperature of the gas, while lower
than in X-ray bright galaxies, is similar to that in other X-ray faint
systems such as NGC~4697 and the bulge of M31 (SIB01; Shirey et al.\ 2001).
The paucity of gas in the system is probably the result of the removal of
most of the gas by supernovae-driven winds rather than interaction with
a cluster environment. The low measured metallicity of the gas is in
conflict with the optically-determined value of the metallicity of the bulge.
The hot gas fills the hole in the H~{\sc i} distribution, possibly
indicating that the hot gas has displaced the cold gas in the overlap
region. However, future observations of Sa galaxies with smaller
H~{\sc i} holes will be able to address this issue more conclusively.

Finally, the central point source is coincident with the optical center of the
galaxy, and exhibits significantly different spectral characteristics than
the sum of the other resolved point sources. Above 1.5 keV, the spectrum
is adequately described by a highly absorbed ($N_H > 10^{22}$ cm$^{-2}$)
power law with $\Gamma = 1.9$, and the luminosity varied by a factor of
two between the two observations. Below 1.5 keV, there is an excess of
soft photons over what is expected from hot gas within the extraction
region. The soft component is of equal strength in the two observations,
and can be modeled by a MEKAL model with a temperature of 0.75 keV
and low metallicity, and has a luminosity that is 5\%--10\% that of the hard
component. The parameters of both the hard and soft components are very
similar to those of low luminosity AGNs studied previously with {\it ASCA}
and {\it ROSAT} (Ptak et al.\ 1999).

\acknowledgments
We thank the referee, Rene Walterbos, for many insightful comments and
suggestions that improved the manuscript considerably.
We thank Jonathan McDowell for his help in implementing the
{\sc calcrmf} and {\sc calcarf} routines of Alexey Vikhlinin for spectral
analysis. We thank David Buote and Elizabeth Blanton for useful
conversations and suggestions.
J. A. I. was supported by {\it Chandra} Fellowship grant PF9-10009,
awarded through the {\it Chandra} Science Center.
C. L. S.  was supported in part by {\it Chandra} Award Numbers
GO0-1141X,
GO0-1158X,
GO0-1173X,
and
GO1-2078X,
all issued by the {\it Chandra} Science Center.
J. N. B. was supported by {\it Chandra} Award Number GO0-1148X.
The {\it Chandra} Science Center is operated by the Smithsonian Astrophysical
Observatory for NASA under contract NAS8-39073.

\end{document}